\documentclass[aps,prd,
floatfix,preprintnumbers,superscriptaddress,nofootinbib,nameinlink,capitalise]{revtex4-2}

\usepackage[left=2.54cm, right=2.54cm, top=2.54cm, bottom=2.54cm]{geometry}
\usepackage{palatino}
\usepackage{amsmath, amssymb}
\usepackage{graphicx}
\usepackage{array, booktabs, multirow}
\usepackage{float}
\usepackage{float} 
\usepackage{hyperref}
\usepackage{subfig}
\usepackage{nameref}      
\usepackage{cleveref}
\usepackage{placeins}
\usepackage{verbatim}
\usepackage{textgreek}
\usepackage[table]{xcolor}
\usepackage{colortbl}
\usepackage[justification=raggedright,singlelinecheck=false]{caption}
\usepackage{color}
\newcommand{\bmat}{\left(\begin{array}}
\newcommand{\emat}{\end{array}\right)}
\newcommand{\be}{\begin{equation}}
\newcommand{\ee}{\end{equation}}
\newcommand{\bea}{\begin{eqnarray}}
\newcommand{\eea}{\end{eqnarray}}

\usepackage{graphicx}
\usepackage{subcaption}

\begin{document}

\title{Revisiting Polynomial Hybrid Inflation: Planck and ACT Compatibility via Radiative Corrections}

\author{Waqas Ahmed}
\email{waqasmit@hbpu.edu.cn}
\affiliation{Center for Fundamental Physics, Hubei Polytechnic University, Huangshi 435003, China}
\affiliation{School of Artificial Intelligence, Hubei Polytechnic University, Huangshi 435003, China}

\author{Saleh O. Allehabi}
\email{s.allehabi@iu.edu.sa}
\affiliation{Department of Physics, Faculty of Science, Islamic University of Madinah, Madinah 42351, Saudi Arabia}

\author{Mansoor Ur Rehman}
\email{mansoor@qau.edu.pk}
\affiliation{Department of Physics, Faculty of Science, Islamic University of Madinah, Madinah 42351, Saudi Arabia}

\noaffiliation


\vspace{1em}
\begin{abstract}
We investigate the impact of one-loop radiative corrections in a non-supersymmetric model of hybrid inflation with a chaotic (polynomial-like) potential,$V(\phi) = V_0 + \lambda_p \phi^p$, in the light of the latest constraints from \textit{Planck} and \textit{Atacama Cosmology Telescope} (ACT) observations. Here, $V_0$ denotes the energy scale of inflation, and $\lambda_p$ is a coupling associated with the polynomial term of power $p$. These corrections can naturally arise from couplings of the inflaton to other matter fields, which also facilitate the reheating process. At the tree level, the predictions of such models for the scalar spectral index $n_s$ and the tensor-to-scalar ratio $r$ typically lie outside the current observational bounds. However, incorporating  one-loop radiative corrections  modifies the potential to,
\[
V(\phi) = V_0 + \lambda_p \phi^p + A \phi^4 \ln (\phi/ \mu),
\]
where $A$ characterizes the strength of the inflaton's coupling to other fields, and \(\mu\) is an appropriate renormalization scale. This radiatively corrected potential can reconcile the model with the combined \textit{Planck}+ACT data over a suitable range of parameter space explored in this work. In particular, radiative corrections from fermionic loops ($A < 0$) suppress the tensor-to-scalar ratio $r$, while simultaneously yielding a red-tilted spectrum with $n_s < 1$, even for sub-Planckian field excursions. This brings the prediction in line with current observations, while still allowing for potentially detectable signatures of primordial gravitational waves.  Furthermore, the inflaton's couplings enable successful reheating and naturally accommodate non-thermal leptogenesis, providing a unified framework for inflation and baryogenesis.
\end{abstract}

\maketitle

\section{Introduction}
The inflationary paradigm offers a compelling explanation for the observed homogeneity, isotropy, and flatness of the universe, as well as the origin of the primordial density fluctuations that seeded the formation of large-scale structure \cite{Guth:1980zm, Linde:1981mu,Albrecht:1982wi}. Among the wide variety of inflationary models, hybrid inflation \cite{Linde:1993cn, Dvali:1994ms, Copeland:1994vg} stands out due to its natural embedding within particle physics frameworks such as grand unified theories, where inflation ends via a symmetry-breaking phase transition. 

In its simplest non-supersymmetric realization, hybrid inflation can feature a chaotic (polynomial-like) inflaton potential of the form
\begin{equation} \label{eq:01}
V(\phi) = V_0 + \lambda_p \phi^p,    
\end{equation}
where $V_0$ sets the inflationary vacuum energy and $\lambda_p$ is a coupling, with mass dimensions $4-p$, associated with the polynomial term of power $p$. However, at the tree level, such models typically predict values for the scalar spectral index $n_s$ and the tensor-to-scalar ratio $r$ that lie outside the bounds set by recent observations from the \textit{Planck} satellite and the Atacama Cosmology Telescope (ACT) \cite{AtacamaCosmologyTelescope:2025nti}.
These predictions can be significantly modified by quantum corrections arising from the inflaton’s couplings to other matter fields—interactions that are also essential for a successful reheating phase. As shown in \cite{Ahmed:2014cma}, incorporating one-loop radiative corrections `a la Coleman-Weinberg \cite{Coleman:1973jx} leads to a modified potential of the form:
\begin{equation} \label{eq:02}
V(\phi) = V_0 + \lambda_p \phi^p + A \phi^4 \ln (\phi/ \mu),    
\end{equation}
where $A$ encodes the strength of the inflaton's coupling to other fields, and \(\mu\) is an appropriate renormalization scale. Radiatively corrected hybrid inflation with $p = 2$, based on Linde’s original model, was first investigated in \cite{Rehman:2009wv} using WMAP5 data \cite{WMAP:2008lyn}, and further extended in \cite{Ahmed:2014cma} for $p = 2/3, 1, 2, 3, 4$, in light of \textit{Planck} 2013 and 2015 results \cite{Planck:2013jfk, Planck:2015sxf}.

In this work, we first consider the radiatively corrected chaotic potential:
\begin{equation}
V(\phi) = \lambda_p \phi^p + A \phi^4 \ln (\phi/ \mu),  
\end{equation}
which was previously studied in \cite{Senoguz:2008nok} for quadratic ($p = 2$) and quartic ($p = 4$) cases. We extend this analysis to include fractional powers $p = 1, 2/3, 1/3$, which are motivated by recent ACT data. Notably, we find that for $p = 1/3$, the model’s predictions can fall within the $1\sigma$ bounds of the most recent ACT constraints, illustrating the potential compatibility of such models with observations.

We then turn to the full radiatively corrected hybrid inflation scenario and provide an updated analysis of its predictions for the polynomial powers $p = 1,2/3,1/3$, in light of the combined data from \textit{Planck}, ACT, and BICEP/Keck 2018 (BK18). Importantly, the case $p = 1/3$ has not been considered in earlier studies. We identify a viable region of parameter space, particularly for $A < 0$, as expected from fermionic loop corrections. These corrections suppress the tensor-to-scalar ratio $r$ and yield a red-tilted scalar spectrum ($n_s < 1$) consistent with current observations. Moreover, they arise naturally from inflaton couplings to heavy right-handed neutrinos, which play a central role in reheating and in generating the observed baryon asymmetry via nonthermal leptogenesis. Such couplings are well motivated in grand unified theories and are an essential ingredient of the type-I seesaw mechanism, which provides a natural explanation for the observed smallness of light neutrino masses. The out-of-equilibrium decays of these nonthermally produced RHNs generate a lepton asymmetry, which is subsequently partially converted into the observed baryon asymmetry via electroweak sphaleron processes.

This work therefore provides the first self-consistent analysis of radiatively corrected hybrid inflation that simultaneously incorporates precision CMB constraints, realistic reheating dynamics, and nonthermal leptogenesis, while remaining compatible with sub-Planckian field excursions and potentially observable primordial gravitational waves \footnote{
Recently, numerous studies have examined both supersymmetric and non-supersymmetric inflationary models in light of observational constraints from the Atacama Cosmology Telescope (ACT); see Refs.~\cite{Rehman:2025fja, Kallosh:2025rni, Zharov:2025zjg, Ketov:2025cqg,Pallis:2025epn, Okada:2025nyd, Pallis:2025nrv,Pallis:2025gii,Pallis:2025vxo, Ellis:2025zrf,Gao:2025onc, Liu:2025qca, Odintsov:2025eiv,Gialamas:2025kef,McDonald:2025tfp, Gialamas:2025kef, Modak:2025bjv,Ahmed:2025eip}.
.}

The structure of the paper is organized as follows. In Section~II, we discuss the radiative corrections to the inflaton potential, outlining the theoretical framework and the motivation for including loop-level effects. Section~III is devoted to the analysis of radiatively corrected chaotic inflation, where we examine how these corrections modify the inflationary dynamics and observational predictions. In Section~IV, we extend our study to radiatively corrected hybrid inflation models that incorporate chaotic potentials, offering a unified framework that connects different inflationary scenarios. Section~V is focused on leptogenesis, where we explore how the inflaton decay can lead to a successful generation of the baryon asymmetry of the Universe. Section~VI presents a comprehensive numerical analysis, identifying the regions of parameter space that are consistent with the most recent cosmological observations from the Planck and ACT collaborations. We conclude in Section~VII with a summary of our findings and a brief discussion of their implications, including possible observational signatures that could test the validity of the proposed framework. 

\section{Radiative Corrections to the Inflaton Potential} \label{sec2}
Considering a generic inflationary setup with one or more scalar fields \(\varphi_i\) driving inflation. The full tree-level action in a curved spacetime background is
\begin{equation}
S = \int d^4x \, \sqrt{-g} \, \left[
\frac{1}{2} \sum_i (\partial_\mu \varphi_i)(\partial^\mu \varphi_i) 
+ \mathcal{L}_{\mathrm{matter}}(\psi_j, \varphi_i) 
- V_{\mathrm{tree}}(\varphi_i, \psi_j) + \cdots
\right],
\end{equation}
The matter Lagrangian \(\mathcal{L}_{\mathrm{matter}}\) may include to couplings of the inflaton (and other scalar fields) to fermionic fields, such as right-handed neutrinos \(N\), or to bosonic fields, such as the SM Higgs boson or GUT Higgs bosons. In realistic inflationary scenarios, the inflaton field generally interacts with other fields present in the theory. These interactions induce quantum corrections to the classical inflaton potential, which can significantly alter the inflationary dynamics and predictions for cosmological observables.

At the one-loop level, the leading quantum corrections are encapsulated by the Coleman-Weinberg effective potential \cite{Coleman:1973jx}. The corrected potential can be expressed as
\begin{equation}
V_{\text{eff}}(\phi) = V_{\text{tree}}(\phi) + \Delta V_{\text{1-loop}}(\phi),
\end{equation}
where \(V_{\text{tree}}(\phi)\) is the tree-level potential and \(\Delta V_{\text{1-loop}}(\phi)\) represents the radiative corrections. The general form of the one-loop contribution arising from fields coupled to \(\phi\) is given by
\begin{equation}
\Delta V_{\text{1-loop}}(\phi) = \sum_i \frac{(-1)^{F_i}}{64\pi^2} M_i^4(\phi) \ln \left( \frac{M_i^2(\phi)}{Q^2} \right),
\end{equation}
where the sum runs over all particle species \(i\), with \(F_i = 0\) for bosons and \(F_i = 1\) for fermions. Here, \(M_i(\phi)\) denotes the field-dependent mass of the \(i\)-th particle and \(Q\) is the renormalization scale.

In many inflationary constructions, the dominant effect can be approximated by a correction of the form
\begin{equation}
\Delta V_{\text{1-loop}}(\phi) = A \, \phi^{4} \ln \left( \frac{\phi}{\mu} \right),
\end{equation}
where the coefficient \(A\) depends on the coupling constants and particle content of the theory, and \(\mu\) is an appropriate renormalization scale. Notably, \(A\) is positive for bosonic loop corrections and negative for fermionic ones. These radiative corrections are crucial for improving the agreement of hybrid and chaotic inflationary models with observational data, such as the scalar spectral index and tensor-to-scalar ratio. Moreover, they provide a natural mechanism linking inflationary dynamics with reheating and baryogenesis through the inflaton’s couplings to other matter fields.

\section{Radiatively Corrected Chaotic Inflation}\label{sec3}

Before we study the effect of radiative correction in hybrid inflation, we first consider a simpler example of radiatively
corrected chaotic inflation. At tree-level, chaotic inflation models are characterized by simple monomial potentials of the form
\begin{equation}
V(\phi) = \lambda_p \, \phi^p,
\end{equation}
where \(\phi\) is the inflaton field and \(p\) is a positive real number that determines the steepness of the potential. These models are appealing due to their simplicity, minimal initial conditions, and predictive power. However, for integer values of \(p\), they are increasingly disfavored by recent observations of the cosmic microwave background (CMB), which constrain the scalar spectral index \(n_s\) and the tensor-to-scalar ratio \(r\) within narrow bounds.

\begin{figure}[ht]
    \centering
    \includegraphics[width=0.48\linewidth,height=6.7cm]{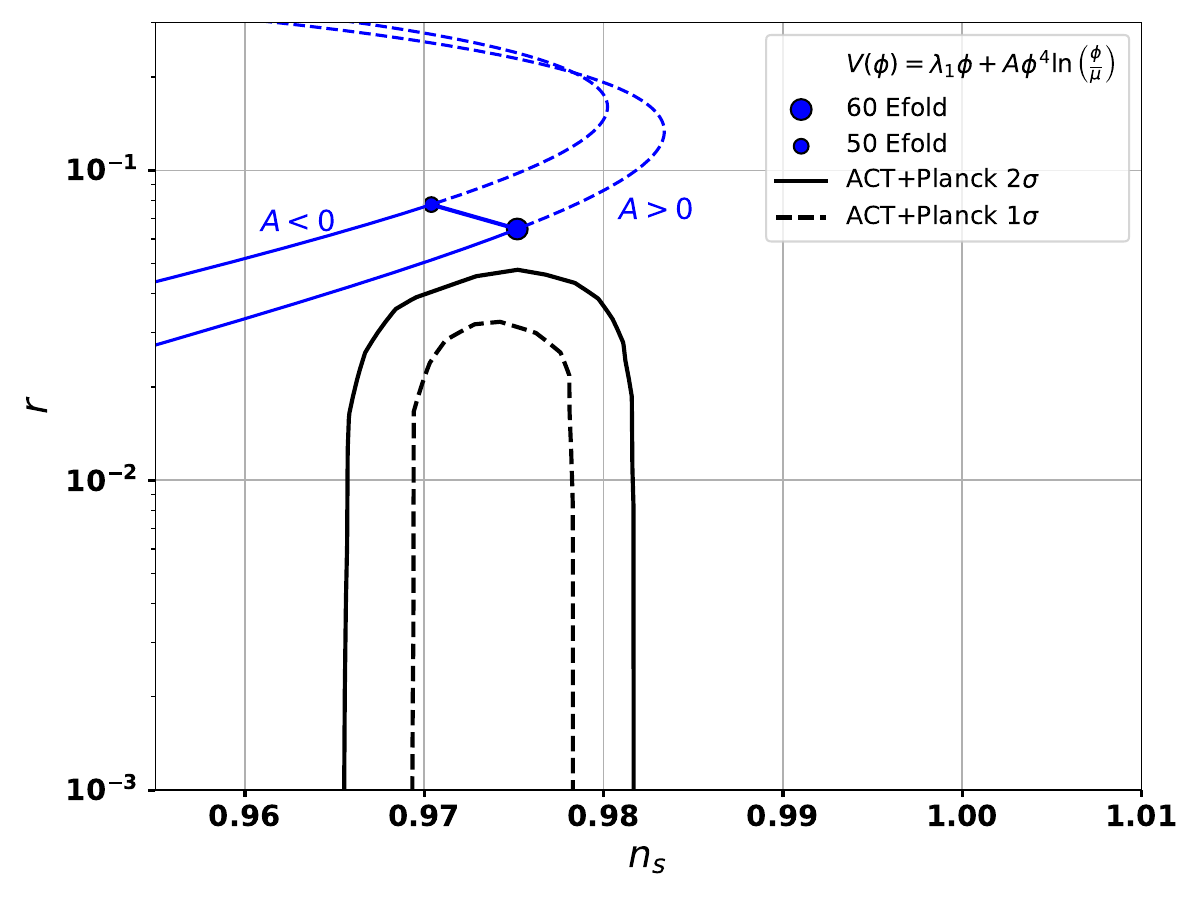}
    \quad
    \includegraphics[width=0.48\linewidth,height=6.7cm]{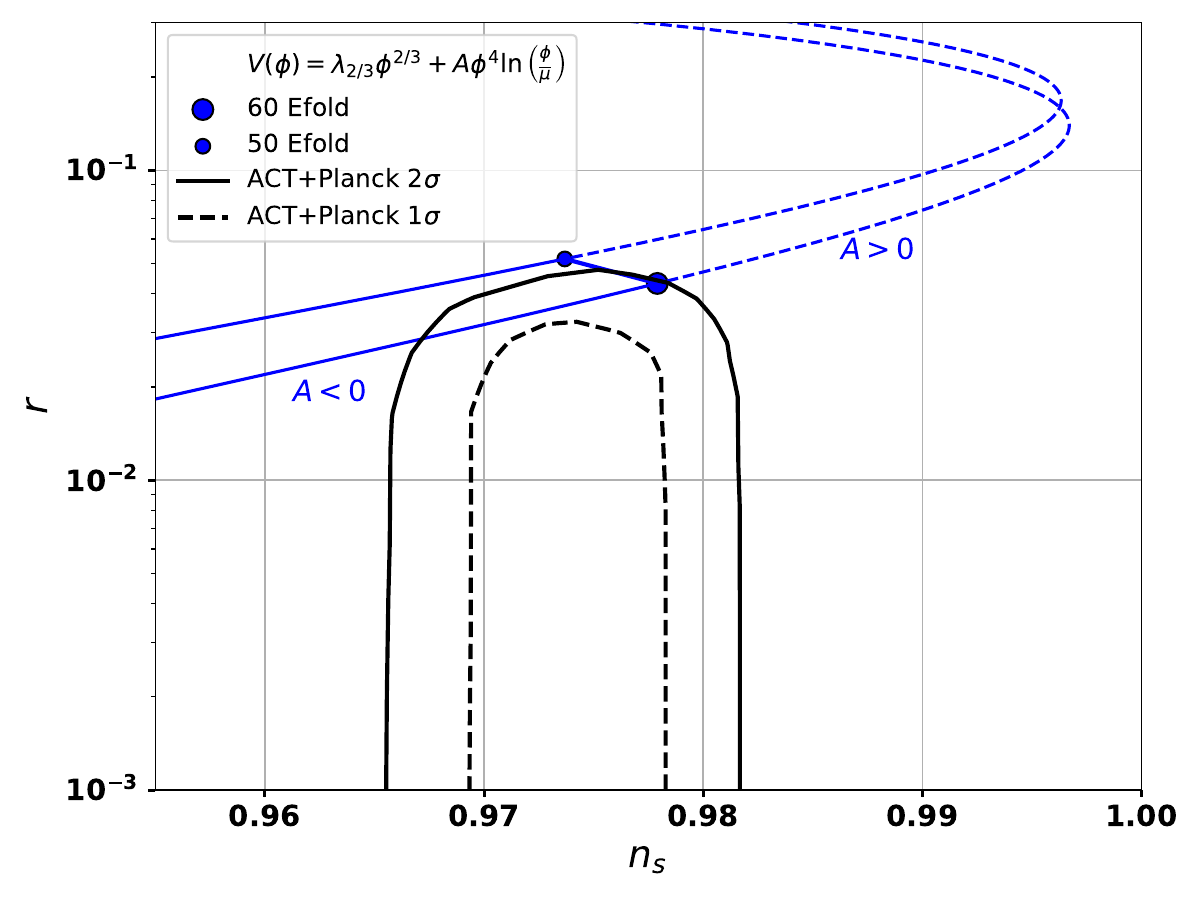}
    \quad
    \includegraphics[width=0.48\linewidth,height=6.7cm]{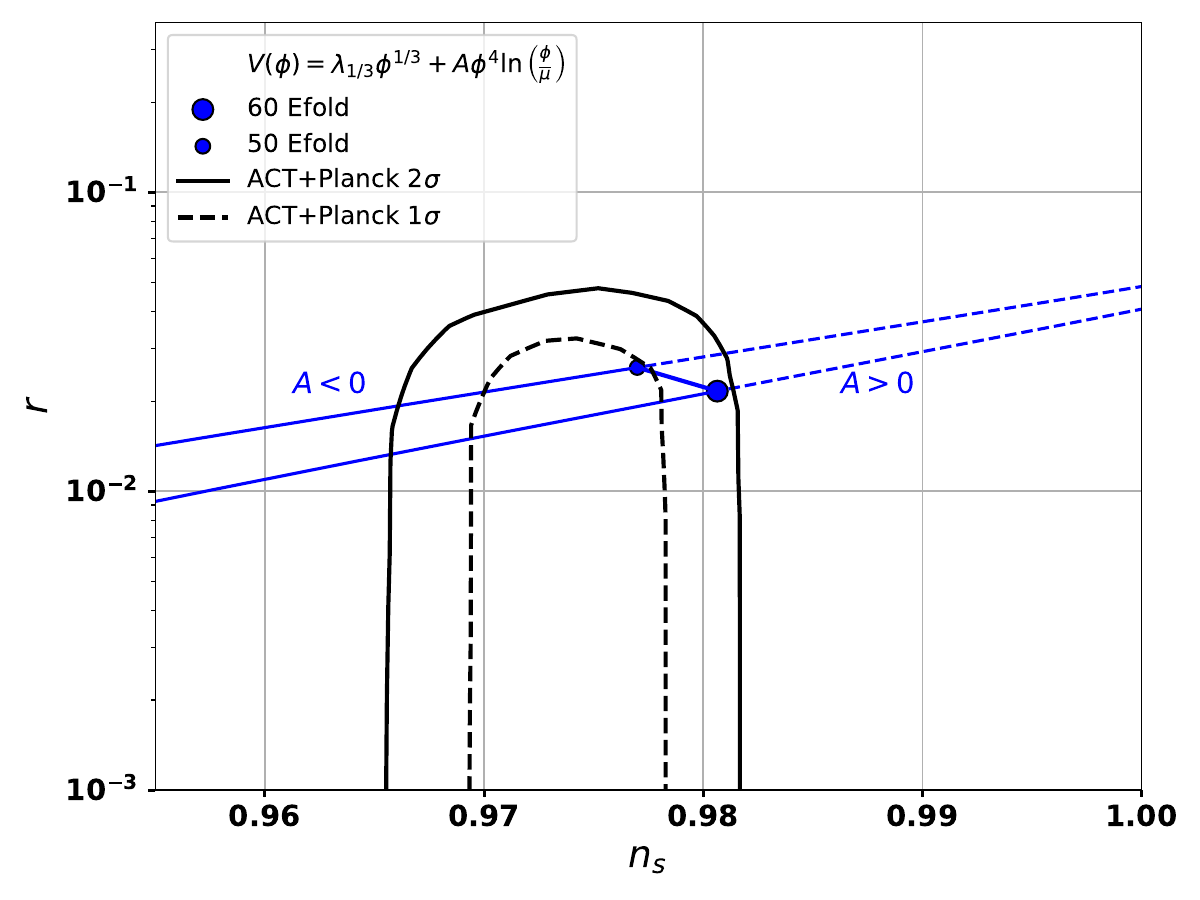}
    \caption{The tensor-to-scalar ratio $r$ is plotted against the scalar spectral index $n_s$, with the black dashed and solid contours indicating the $1\sigma$ and $2\sigma$ confidence regions derived from the combined Planck 2018, ACT DR6, and BICEP/Keck 2018 datasets (P+ACT+LB+BK18)~\cite{AtacamaCosmologyTelescope:2025nti}. The small and large blue dots correspond to predictions for $N = 50$ and $N = 60$ $e$-folds, respectively. The solid lines represent predictions from the radiatively corrected potential with fermionic loops ($A<0$), whereas the dashed lines correspond to the bosonic correction scenario ($A>0$).}
    \label{fig1}
\end{figure}

Using the standard slow-roll approximations, as defined in the appendix, the tree-level predictions of the scalar spectral index \(n_s\), the tensor-to-scalar ratio \(r\), and the running of the scalar spectral index, $\alpha_s$ are given by:
\begin{equation}
n_s \approx 1 - \frac{p + 2}{2N_0}, \quad r \approx \frac{4p}{N_0}, \quad
\alpha_s \equiv \frac{d n_s}{d \ln k} \approx \frac{p^2}{16N_0^2} \left[ -24 + \frac{4(p-1)(10 - p)}{p^2} \right]
\label{eq:ns_r_treelevel}
\end{equation}
where \(N_0\) is the number of $e$-folds before the end of inflation. The predictions for $n_s$ and $r$ are illustrated in Fig.~\ref{fig1} by the blue dots for representative values of \(N = 50\) and \(N_0 = 60\). The running of the scalar spectral index, $|\alpha_s| < 10^{-3}$, is consistent with the current observations.

Quantum corrections arising from the inflaton’s interactions with other fields modify the classical potential. Incorporating the leading one-loop radiative correction following the Coleman-Weinberg mechanism~\cite{Coleman:1973jx}, the effective potential takes the form
\begin{equation}
V(\phi) = \lambda_p \, \phi^p + A \, \phi^4 \ln \left( \frac{\phi}{\mu} \right),
\label{eq:rad_corrected_potential}
\end{equation}
where \(A\) encodes the loop-level contributions and depends on the couplings of the inflaton to other matter fields, and \(\mu\) is the renormalization scale. The second term can arise from either bosonic or fermionic quantum loops, with the sign and magnitude of \(A\) determined by the spin and statistics of the particles coupled to the inflaton.

Radiative corrections can substantially alter the inflationary dynamics and predictions. The logarithmic term in Eq.~\eqref{eq:rad_corrected_potential}, induced by quantum loops, modifies both the scalar spectral index \(n_s\) and the tensor-to-scalar ratio \(r\). As shown in Fig.~\ref{fig1}, for fermionic corrections (\(A < 0\)), both observables shift to lower values, typically yielding a redder spectrum and suppressed tensor modes, thereby improving compatibility with CMB constraints. In contrast, bosonic corrections (\(A > 0\)) increase both \(n_s\) and \(r\), often pushing predictions outside the preferred observational bounds. Using the latest combined Planck 2018, ACT DR6, and BICEP/Keck 2018 data (P+ACT+LB+BK18)~\cite{AtacamaCosmologyTelescope:2025nti}, we find that for fermionic corrections, the model with \(p = 1/3\) falls well within the $1\sigma$ region for both \(N = 50\) and \(N = 60\), while \(p = 2/3\) is allowed at \(2\sigma\) for \(N = 60\). However, the linear case \(p = 1\) lies outside even the $2\sigma$ region. In contrast, bosonic corrections displace all three models, \(p = 1/3\), \(2/3\), and \(1\), beyond the $2\sigma$ contours, rendering them incompatible with the data.

To illustrate the analytic behavior, consider the following radiative corrections into otherwise tree-level expressions:
\begin{equation}
n_s \approx 1 - \frac{p + 2}{2 N_0} + \frac{A (2 p N_0)^{\frac{4-p}{2}}}{2 \lambda_p N_0} \, \mathcal{F}_1(p, \ln(\phi_*/\mu)), \quad r  \approx \frac{4 p}{N_0} \left[ 1 + \frac{2 A (2 p N_0)^{\frac{4-p}{2}}}{p \lambda_p} \left( 4 \ln \frac{\sqrt{2 p N_0}}{\mu} + 1 \right) \right],
\end{equation}
where
\begin{equation}
\mathcal{F}_1 = 12 \ln \left( \frac{\sqrt{2 p N_0}}{\mu} \right) + 7 - \frac{3 p \left( 4 \ln \left( \frac{\sqrt{2 p N_0}}{\mu} \right) + 1 \right)}{p + 2}.
\end{equation}
For example, choosing \(p = 1/3\) and \(N = 60\) yields \(n_s \approx 0.973\) and \(r \approx 0.016\), which lie well within the $1-\sigma$ observational region. In short, radiatively corrected polynomial inflation models with fractional powers, such as \(p = 2/3\) and \(p = 1/3\), have been shown to yield better fits to the current observational data compared to models with power greater than unity like linear (\(p=1\)) or quadratic (\(p=2\)) inflation. These fractional models with $p < N_0/100$ naturally predict a suppressed tensor-to-scalar ratio $r < 0.04$ and a red-tilted spectrum ($n_s < 1-1/N_0$), bringing them into good agreement with current observations. To alleviate the tension between chaotic inflation models with polynomial powers $p > N_0/100$ and observational constraints, we consider a hybrid inflation framework, as detailed in the next section. 
This framework effectively introduces an additional parameter, $V_0$, into the single-field inflationary potential. An alternative and equally compelling possibility involves introducing a non-minimal coupling between the inflaton and gravity through a term like $\xi \phi^2 \mathcal{R}$, where $\mathcal{R}$ is the Ricci scalar and $\xi$ is a dimensionless coupling constant. For a quartic potential with bosonic radiative corrections, such models can yield predictions consistent with recent ACT data, as demonstrated in \cite{Ahmed:2025rrg}, while also forecasting potentially observable gravitational wave signatures within reach of future CMB polarization experiments \cite{LiteBIRD:2022cnt, CMB-S4:2020lpa, SimonsObservatory:2018koc}. For related work, see also \cite{Bostan:2019fvk}.

\section{Radiatively Corrected Hybrid Inflation with Chaotic Potentials} \label{sec4}
We consider a non-supersymmetric hybrid inflation model characterized by a real scalar inflaton field \(\phi\) and a real waterfall field \(\chi\). The hybrid inflation (HI) potential is composed of three main components: the Higgs potential of the waterfall field \(V(\chi)\), the self-interaction potential of the inflaton \(\delta V(\phi)\), and an interaction term proportional to \(g^2 \chi^2 \phi^2\) that couples the two fields. Together, the tree-level hybrid inflation (TLHI) potential is given by
\begin{equation}
V(\phi, \chi) = V(\chi) + \delta V(\phi) + \frac{g^{2}}{4} \phi^{2} \chi^{2}.
\end{equation}
where
\begin{equation}
V(\chi) = \kappa^2 \left( M^2 - \frac{\chi^2}{4} \right)^2, \quad  \delta V(\phi) = \lambda_p \phi^p .    
\end{equation}
Here, $ \kappa, g > 0$ are real value parameters, and $M$ denotes the symmetry-breaking scale associated with the waterfall field. The quadratic potential with $p=2$ and $\lambda_2 = \frac{1}{2} m^2$ corresponds to Linde's original hybrid inflation model \cite{Linde:1993cn}.
The inflationary dynamics unfold along the $\chi = 0$ trajectory. The effective mass squared of $\chi$ in this direction is given by 
\begin{equation}
    m_\chi^2(\phi) = -\kappa^2 M^2 + \frac{g^2}{2} \phi^2
\end{equation}
For field values sufficiently larger than the critical value, $\phi_c \equiv \frac{\sqrt{2} \kappa M}{g}$, the mass squared remains positive and the $\chi$ field is stabilized at the origin. In this regime, the dynamics are effectively governed by a single-field potential of the form
\begin{equation}
V(\phi) = V_0 + \lambda_p \phi^p, \quad \text{with} \quad V_0 = \kappa^2 M^4,
\end{equation}
where the false vacuum energy $V_0$ dominates the potential and drives a quasi-de Sitter phase of exponential expansion. Inflation persists until the inflaton field slowly rolls down to the critical value $\phi = \phi_c$. At this point, the effective mass of $\chi$ becomes tachyonic, destabilizing the $\chi = 0$ configuration. The field $\chi$ then rapidly acquires a nonzero VEV, triggering a spontaneous symmetry-breaking phase transition—often referred to as a ``waterfall'' transition—which abruptly terminates inflation. This graceful exit mechanism is a central feature of hybrid models and allows for a natural end to the inflationary epoch. 

Assuming $V_0 \gg \lambda_p \phi^p$ for $p \leq 1$, the tree-level slow-roll predictions for the scalar spectral index and the tensor-to-scalar ratio are given by
\begin{equation}
n_s \approx 1 - \frac{2(p - 1)}{2 - p} \cdot \frac{1}{N_0}, \quad r \approx \frac{8p^2}{(2 - p)^2} \cdot \frac{1}{N_0^2}.
\end{equation}
For $p \leq 1$, this results in a blue-tilted scalar spectrum ($n_s \geq 1$) and a suppressed tensor-to-scalar ratio ($r \lesssim 8/N_0^2$). The running of the spectral index is also negligible, scaling as $|\alpha_s| \propto N_0^{-2}$. These tree-level predictions are inconsistent with current ACT observational data due to the blue tilt in $n_s$. In the following, we examine how radiative corrections can modify these predictions and restore compatibility with observations.

In standard inflationary scenarios, the Universe transitions into a radiation-dominated phase through the decay of the inflaton into lighter particles, a process known as reheating. For this to occur, the inflaton must couple to other fields. In this work, we consider a set of well-motivated interaction terms and explore how these couplings can influence the inflationary dynamics itself. A minimal and natural choice for these fields includes the waterfall Higgs scalar $\chi$ and right-handed neutrinos $N_i$.

The full tree-level action involving the inflaton $\phi$, the waterfall field $\chi$, and the additional field $N$ is given by
\begin{equation}
S \supset \int d^4x \, \sqrt{-g} \, \Bigg[
 - V(\phi, \chi) 
- \frac{1}{2} \left( y_\phi \phi + m_N \right) \overline{N^c} N 
+ \frac{\lambda_{\chi}}{2 \, m_P} \chi^2 \overline{N^c} N  + \cdots \Bigg], 
\end{equation}
with the hybrid scalar potential defined as
\begin{equation}
V(\phi, \chi) =  \kappa^2 \left( M^2 - \frac{\chi^2}{4} \right)^2 + \frac{g^{2}}{4} \phi^{2} \chi^{2} +  \lambda_p \phi^p.   
\end{equation}
The term $m_N$ denotes the bare Majorana mass of the right-handed neutrinos, while the Yukawa coupling $y_\phi$ introduces a field-dependent mass term of the form $(y_\phi \phi + m_N)\, \overline{N^c} N$. This interaction not only plays a crucial role in realizing the seesaw mechanism for generating light neutrino masses, but also facilitates inflaton decay and subsequent reheating of the Universe.

During inflation, while all other fields are assumed to be stabilized at the origin, they still contribute indirectly through radiative corrections to the inflaton potential. These loop-induced corrections arise due to vacuum fluctuations of the waterfall Higgs field $\chi$ and the right-handed neutrinos $N$ in the inflationary background. Such effects can be systematically captured using the Coleman–Weinberg one-loop effective potential, which for the inflaton $\phi$ takes the form
\begin{align}
V_{\rm eff}(\phi) \simeq \, & V_{0} + \lambda_p \phi^p + \frac{1}{64 \pi^2} \Bigg[ \left(p(p-1) \lambda_p \phi^{(p-2)} \right)^2  \ln\left( \frac{ p(p-1) \lambda_p \phi^{(p-2)} }{\mu^2} \right) + \frac{g^4}{4} \left(\phi^2 - \phi_c^2 \right)^2 \ln \left( \frac{\frac{g^2}{2} (\phi^2 - \phi_c^2)}{\mu^2} \right)  \nonumber \\
& - 2 \mathcal{N}_N \left(y_\phi \phi + m_N \right)^4 \ln \left( \frac{\left( y_\phi \phi + m_N \right)^2}{\mu^2} \right) + \cdots
\Bigg],
\end{align}
where $\mathcal{N}_N = 3$ denotes the number of right-handed neutrino species. For fermionic corrections to be dominant, we assume $g < (6)^{1/4} y_{\phi}$.

In the inflationary regime where the field values are large, specifically, when $\lambda_{\phi H} \phi^2 \gg m_H^2$ and $y_\phi \phi \gg m_N$, the effective potential simplifies to
\begin{equation} \label{eq:Veff}
V_{\rm eff}(\phi) \simeq V_{0} + \lambda_p \phi^p + \phi^4 \left[ \kappa_b \ln \left( \frac{\phi}{\mu_b} \right) - \kappa_f \ln \left( \frac{\phi}{\mu_f} \right) \right] = V_0 + \lambda_p \phi^p + A \phi^4 \ln \left( \frac{\phi}{\mu} \right),
\end{equation}
where the bosonic and fermionic loop coefficients are given by
\begin{equation}
\kappa_b = \frac{1}{16 \pi^2} \frac{g^4}{8}, \quad
\kappa_f = \frac{\mathcal{N}_N}{16 \pi^2} y_\phi^4,
\end{equation}
and the renormalization scales are defined as $\mu_b = \mu / (g/\sqrt{2})$ and $\mu_f = \mu / y_\phi$. For simplicity, we introduce a common loop coefficient defined as, $A \equiv \max(\kappa_b, \, \kappa_f) $, which allows us to express the logarithmic correction in a unified way.. Typically, $A > 0$ corresponds to a boson-dominated loop correction, while $A < 0$ indicates a fermion-dominated scenario.

These radiative corrections can significantly modify the inflaton potential, particularly in the large-field regime. Notably, when $A < 0$, the logarithmic term flattens the potential, leading to a suppressed tensor-to-scalar ratio and a red-tilted spectral index ($n_s < 1$). This behavior improves the model’s agreement with current CMB observations, including recent results from the Planck mission and the Atacama Cosmology Telescope (ACT).
In the following, we explore the inflationary dynamics with these loop-induced corrections and compute the key observables that characterize the inflationary phase: the scalar spectral index $n_s$, the tensor-to-scalar ratio $r$, and the amplitude of scalar perturbations. After discussing the details of reheating, we confront the model’s predictions with the latest observational constraints.

\section{Reheating and Leptogensis} \label{sec5}
It is important to emphasize that the purpose of the present work is to highlight the role of quantum corrections and reheating consistency in inflationary models with mostly fractional-power polynomial potentials. For $p \leq 1$, the simple tree-level potential is not well behaved near the origin and does not possess a true vacuum. We therefore assume that the effective potential is modified after inflation by additional terms that are negligible during inflation but become relevant near the minimum, allowing for a consistent reheating phase. A representative example is obtained by supplementing the potential with an inverse-power term ($\sim 1/\phi^{p}$) as follows:
\begin{equation}
V^{\text{tree}} (\phi) \simeq \lambda_p \phi^p + \frac{\kappa_p }{\phi^p},
\end{equation}
for $p \leq 1$. The inclusion of second term stabilizes the potential and generates a nonzero vacuum expectation value for the inflation, $\langle \phi \rangle$, given by
\begin{equation}
\langle \phi \rangle \simeq \left( \frac{\kappa_p }{\lambda_p} \right)^{1/2p}.
\end{equation}
By choosing $\kappa_p$ sufficiently small, we ensure $\langle \phi \rangle \ll m_P$, so that this additional term is strongly suppressed during inflation ($\phi \lesssim m_P$) and does not affect the inflationary dynamics or the predictions presented in this work. Instead, it becomes relevant only during the post-inflationary phase, allowing the inflaton–waterfall field system to oscillate around a well-defined vacuum and reheat the Universe consistently.
Such modifications are well motivated within effective field theory and may arise from ultraviolet physics, including string-inspired constructions. As an explicit example, a concrete realization of this mechanism for $p=2/3$ has recently been discussed in Ref. \cite{Hai:2025wvs}. 

The radiation content of the early universe is generated by the decay of the inflaton field \( \phi \) after the end of inflation. This process converts the inflaton’s vacuum energy into a thermal bath of relativistic particles, initiating the reheating era. To estimate the reheating temperature \( T_{\rm r} \), we compute the total decay width \( \Gamma_{\text{inf}} \), accounting for all kinematically allowed decay channels into scalar and fermionic degrees of freedom.

In the hybrid inflation model under consideration, the oscillating system \( \phi,\, \chi \) decays into right-handed neutrinos $NN$ as follows:
\begin{itemize}

\item \textbf{Inflaton scalar} $\phi$ decays via the Yukawa interaction $y_{\phi} \phi NN$:
\begin{equation}
\mathcal{L}_{\phi NN} = -\frac{1}{2} y_\phi \phi \overline{N^c} N + \text{h.c.}
\end{equation}
The decay width is given by:
\begin{equation}
\Gamma_{\phi \to NN} = \frac{y_\phi^2 m_\phi}{8\pi} \left(1 - \frac{4 M_N^2}{m_\phi^2} \right)^{3/2},
\end{equation}
where the effective Majorana mass of the right-handed neutrino and the inflaton mass read as.
\begin{eqnarray}
M_N = y_{\chi} M + m_N, \quad   m_\phi \simeq \sqrt{2} g M, 
\end{eqnarray}
where $y_{\chi} \equiv 4 \lambda_{\chi} M/m_P$, and $\langle \phi \rangle  \simeq 0$ and $\langle \chi \rangle =2M$ has been used.

\item \textbf{Waterfall scalar} \( \chi \), decays via the coupling \( \frac{\lambda_{\chi}}{2m_P} \chi^2 N N \). The decay width is:
\begin{equation}
\Gamma_{\chi \to N N} = \frac{y_\chi^2 m_\chi}{8\pi} \left(1 - \frac{4 M_N^2}{m_\chi^2} \right)^{3/2},
\end{equation}
provided \( m_\chi \simeq \sqrt{2} \kappa M  > 2 m_N \).
\end{itemize}

\noindent
The total decay width of the inflaton is then:
\begin{equation}
\Gamma_{\text{inf}} = \Gamma_{\phi \to NN} + \Gamma_{\chi \to NN} .
\end{equation}

\noindent
Assuming rapid thermalization, the reheating temperature is given by:
\begin{equation}
T_{\rm r} \approx \left( \frac{90}{\pi^2 g_*} \right)^{1/4} \sqrt{\Gamma_{\text{inf}} m_P},
\end{equation}
where \( g_* = 106.75 \) is the effective number of relativistic degrees of freedom at reheating.

For a recent discussion of non-thermal leptogenesis in the SM with right-handed neutrinos, see \cite{Okada:2025daq}. The ratio of the lepton number density to the entropy density in the limit $T_r < M_N \equiv M_{N_1}\leq m_{\text{inf}} /2 < M_{N_{2,3}}$ is defined as
\begin{equation} \label{nLs}
\frac{n_{L}}{s} \simeq \frac{3}{2} \frac{T_{r}}{m_{\phi}}\epsilon_{cp}\,,
\end{equation}
where $\epsilon_{cp}$ is the CP-asymmetry factor, which is generated from the out-of-equilibrium decay of the lightest right-handed neutrino. For a normal hierarchical pattern of the light neutrino masses, this factor becomes \cite{Okada:2025daq} 
\begin{equation}
\epsilon_{cp} \simeq \frac{3}{8\pi}\frac{M_{N} m_{\nu_{3}}}{v_{u}^2}\delta_{\rm eff}\,, 
\end{equation}
where $m_{\nu_3}$ is the mass of the heaviest light neutrino, $v_{h}$ is the VEV of the SM Higgs, and $\delta_{\rm eff}$ is the CP-violating phase. 
In the numerical estimates discussed below, we take $m_{\nu_3} = 0.05$ eV, $|\delta_{\rm eff}|\le1$, $v_h = 246$ GeV. A successful baryogenesis is usually generated through the sphaleron process \cite{tHooft:1976rip,Manton:1983nd,Klinkhamer:1984di}, where an initial lepton asymmetry is partially converted into baryon asymmetry expressed as $n_{B}/s=-0.35 \, n_L/s$ \cite{Kuzmin:1985mm,Arnold:1987mh,Khlebnikov:1988sr}.
This framework establishes a quantitative link between the inflaton decay channels and the onset of the radiation-dominated era, enabling scenarios such as non-thermal leptogenesis via the out-of-equilibrium decays of heavy right-handed neutrinos.

\section{Numerical Analytics} \label{sec6}
To identify regions of parameter space consistent with current observational data, we perform a systematic random scan over the following set of fundamental parameters:
\begin{eqnarray}
&& 10^{15} < M < m_P, \quad M < \phi_0 \leq  m_P,  \quad 10^{12} < \lambda_p^{1/(4-p)}/\text{GeV} \leq 10^{16}, \\    
&& 10^{-7} \leq \kappa = g \leq \sqrt{4 \pi}, \quad  10^{-7} \leq  y_{\phi} = y_{\chi} \leq \sqrt{4 \pi}, \quad  10^{-17} \leq  |A| \leq 10^{-13}.
\end{eqnarray}
The effective scalar potential, given in Eq.~\eqref{eq:Veff}, primarily depends on three key parameters: the dimensionless coupling $\lambda_p$, the vacuum energy $V_0 = \kappa^2 M^2$, and the radiative correction coefficient $A$. In addition, the model includes the Yukawa couplings $y_\phi$ and $y_\chi$, which are central to the reheating phase, allowing the inflaton to decay into right-handed neutrinos. These couplings naturally link the inflationary dynamics to the origin of the baryon asymmetry via non-thermal leptogenesis. For the purpose of our numerical computations, we fix the renormalization scale at the critical field value, i.e.,  $\mu = \phi_c$.

It is important to note that when fermionic radiative corrections dominate ($A<0$), the effective potential becomes unbounded from below at large field values, which signals a breakdown of the low-energy effective description rather than a physical inconsistency. We therefore restrict our analysis to sub-Planckian inflaton field values, where slow-roll inflation takes place and the potential remains well behaved. We assume that at trans-Planckian field values the potential is stabilized by ultraviolet physics through higher-dimensional operators with positive coefficients, which do not spoil the inflationary predictions in the regime of interest.

We apply the following observational constraints, which any viable solution must satisfy:
\begin{itemize}
\item The amplitude of the scalar power spectrum, with the slow-roll expression derived in the appendix, is measured to be
\begin{equation}
A_s (k_0) = 2.137 \times 10^{-9},
\end{equation}
at the pivot scale \(k_0 = 0.05\, \mathrm{Mpc}^{-1}\)\cite{Planck:2018vyg}.  
\item   The total number of e-folds $N_0$ connecting inflation to the thermal history of the Universe, under standard reheating assumptions, is approximated by~\cite{1990eaun.book.....K, Liddle:2003as}:
\begin{equation}
    N_0 \simeq 53 + \frac{1}{3} \ln \left( \frac{T_r}{10^{9}\, \mathrm{GeV}} \right) + \frac{2}{3} \ln \left( \frac{V_0}{10^{15}\, \mathrm{GeV}} \right).
\label{eq:efolds}
\end{equation}
The inflaton decay and reheating details have been discussed in the preceding section.
\item The observed baryon-to-entropy ratio, related to the lepton asymmetry via electroweak sphalerons ($n_B/s = -0.35 , n_L/s$), is given by \cite{Planck:2018vyg}:
\begin{equation}
n_{B}/s   = (8.2 - 9.2) \times 10^{-11} \, , 
\end{equation}
where $n_L/s$ is defined in Eq.~\ref{nLs}.
\end{itemize}

\begin{figure}[htbp]
    \centering
    \includegraphics[width=0.48\linewidth,height=6.7cm]{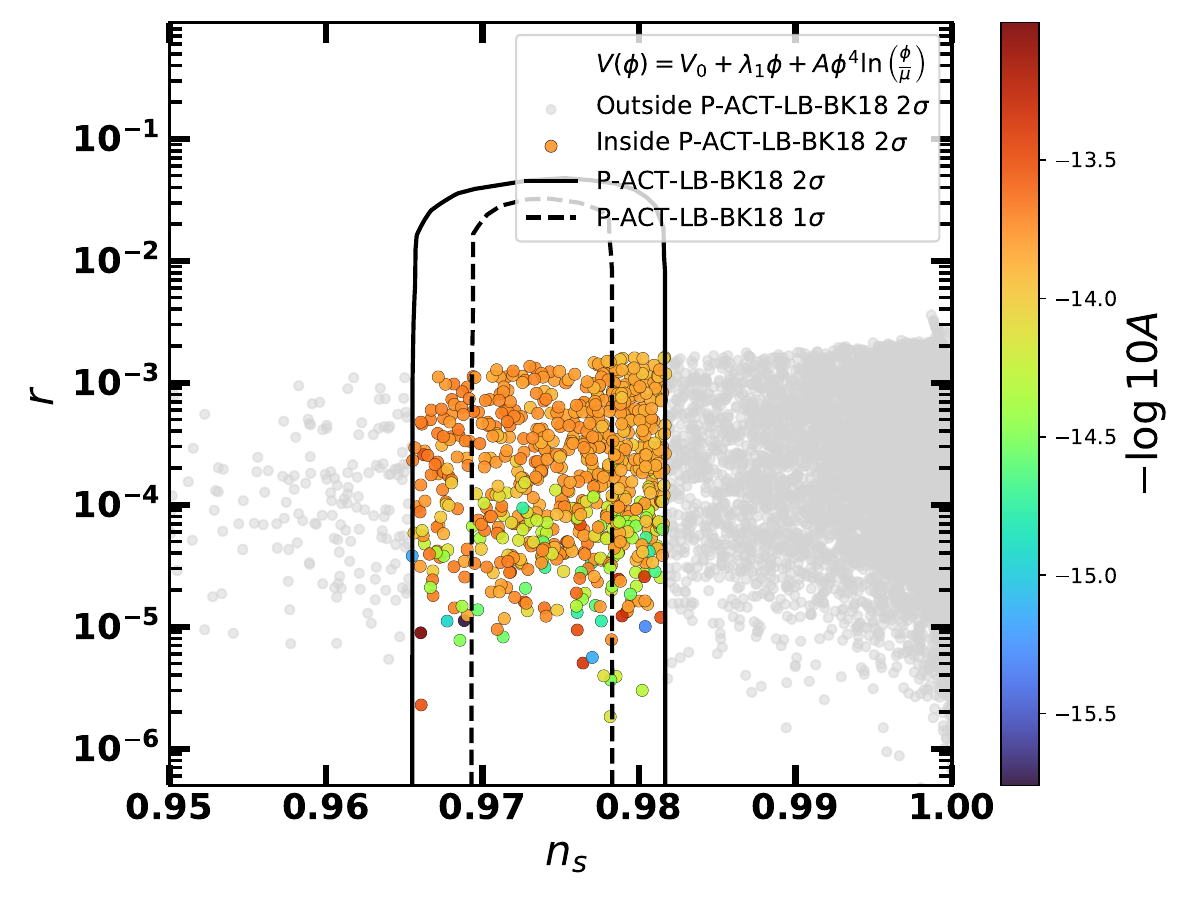}
    \quad
        \includegraphics[width=0.48\linewidth,height=6.7cm]{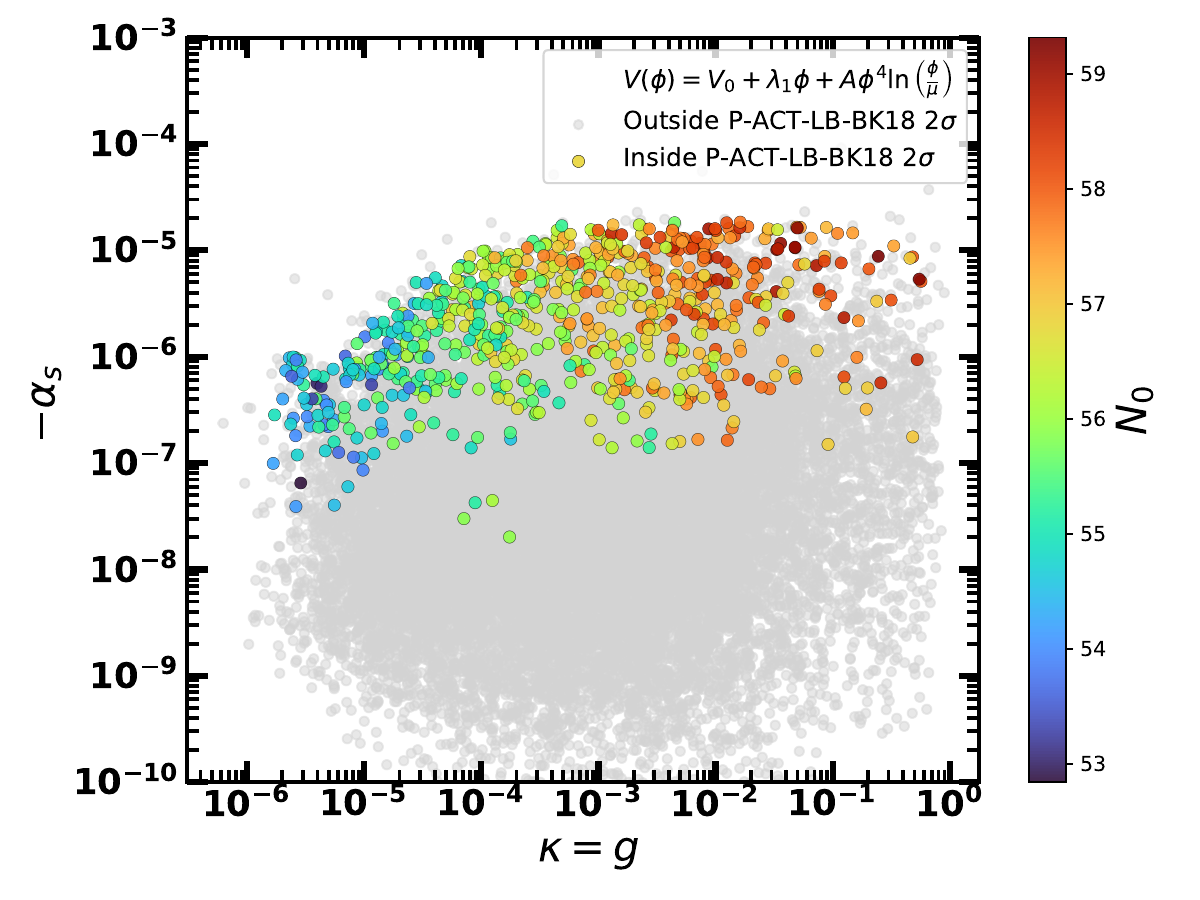}
    \quad
    \includegraphics[width=0.48\linewidth,height=6.7cm]{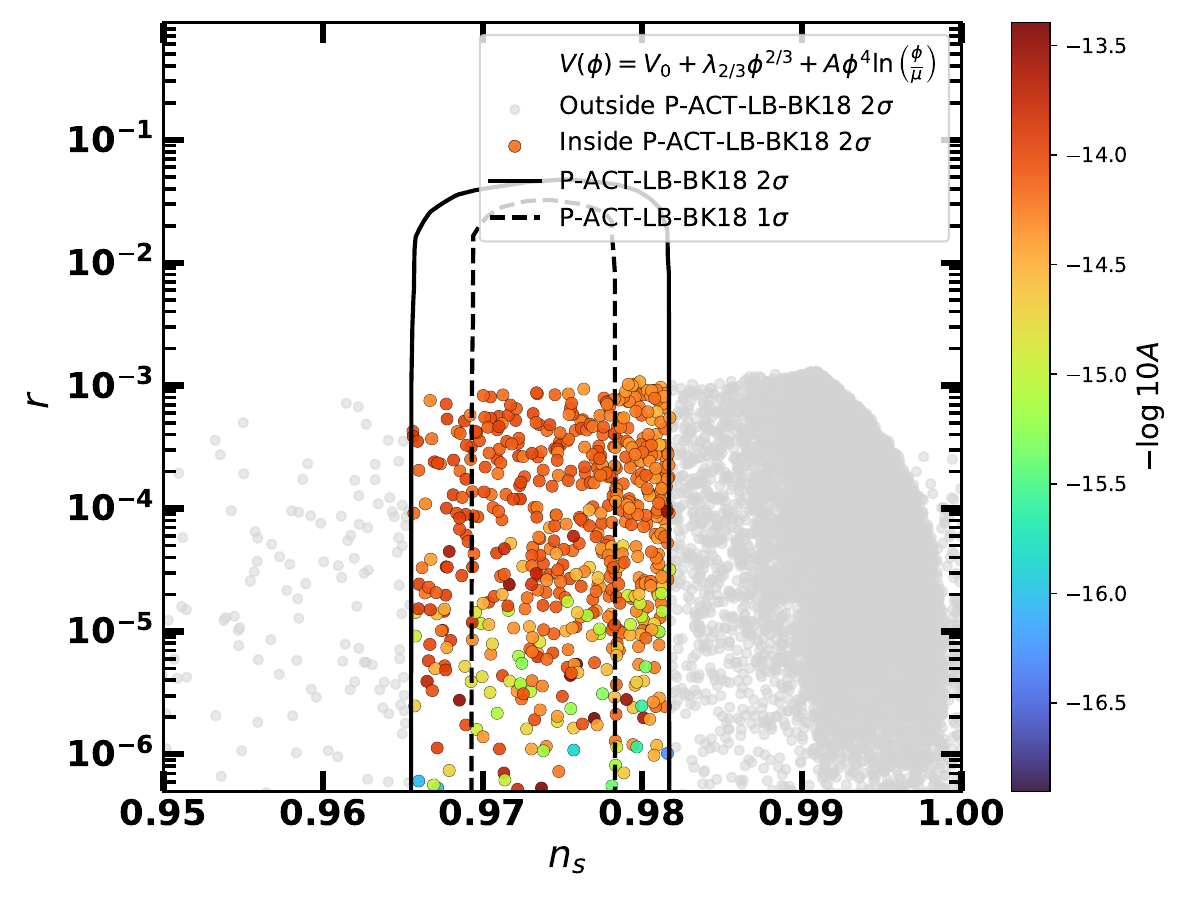}
      \quad
          \includegraphics[width=0.48\linewidth,height=6.7cm]{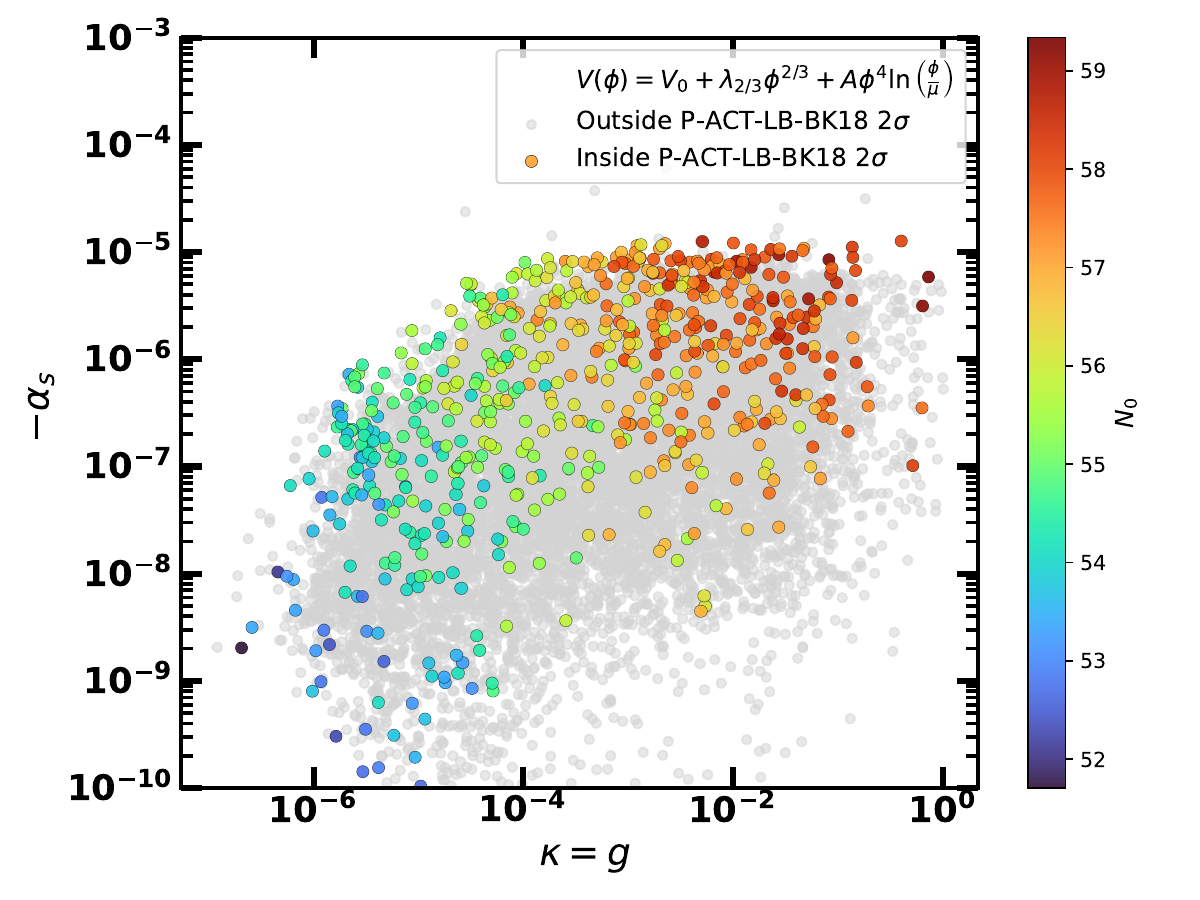}
    \quad
      \includegraphics[width=0.48\linewidth,height=6.7cm]{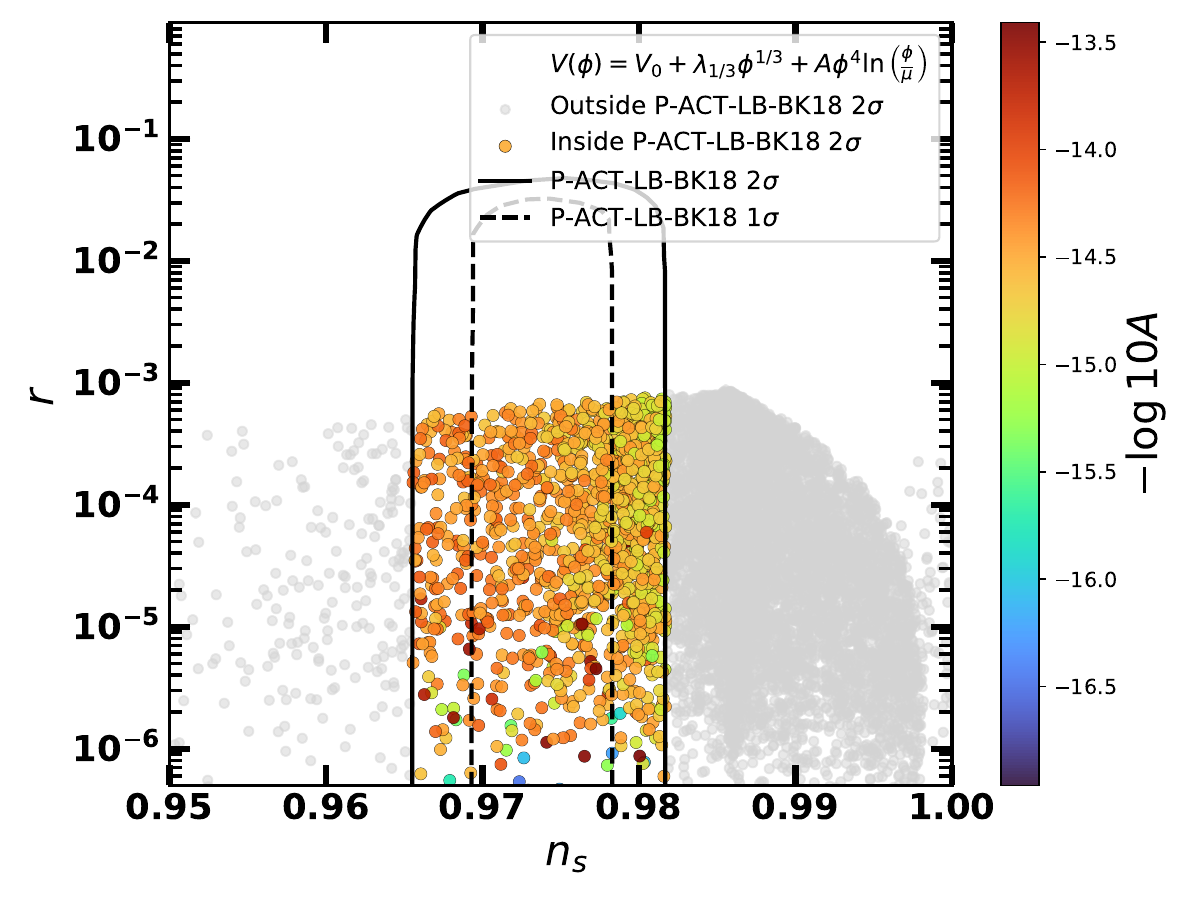}
      \quad
          \includegraphics[width=0.48\linewidth,height=6.7cm]{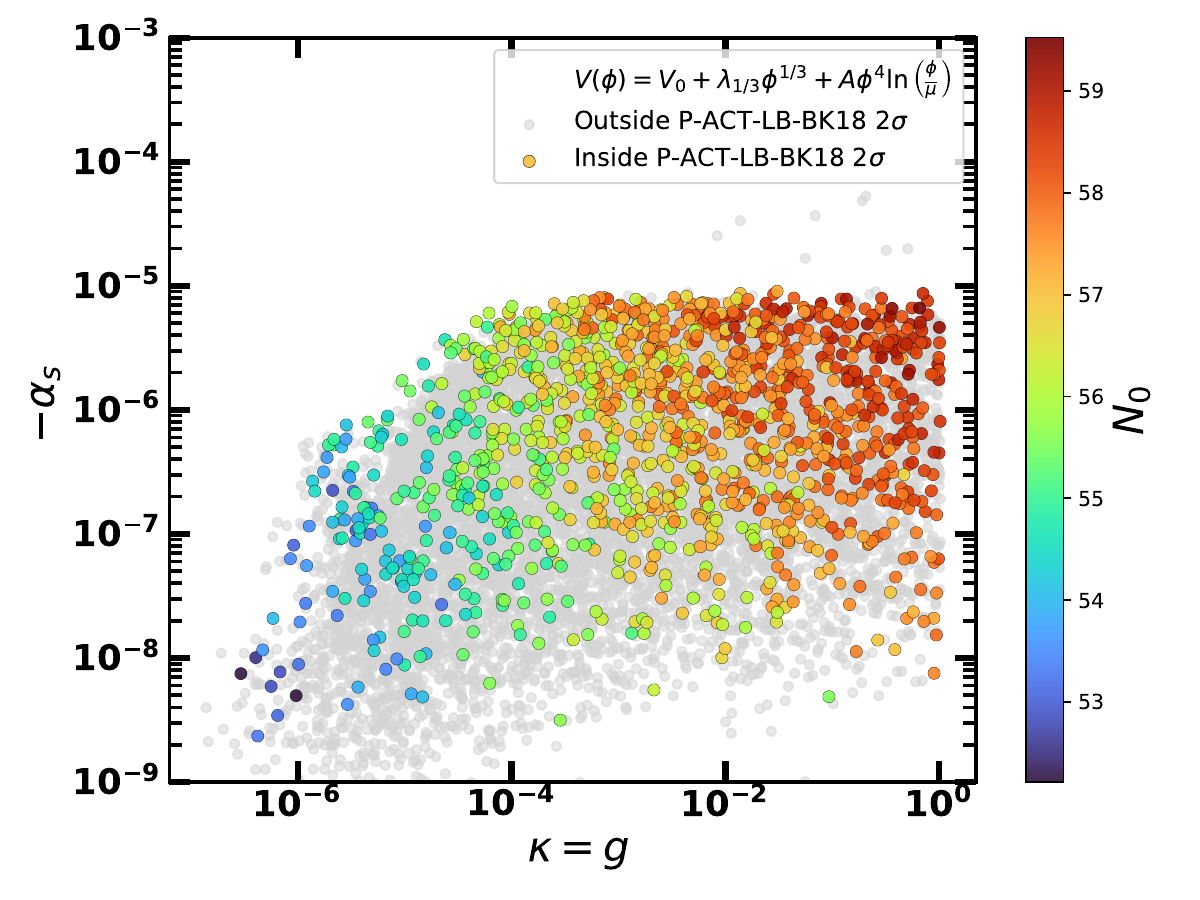}
    \quad
     
      \quad
     \caption{
The left panel illustrates the tensor-to-scalar ratio \( r \) versus the scalar spectral index \( n_s \), where the color coding represents the radiative correction parameter \( A \), showcasing its influence on inflationary predictions. Overlaid are the 1\(\sigma\) (dotted) and 2\(\sigma\) (dashed) confidence contours derived from the combined datasets of Planck 2018, ACT DR6, and BICEP/Keck 2018 (P + ACT + LB + BK18)~\cite{AtacamaCosmologyTelescope:2025nti}. The right panel shows the running of the scalar spectral index \( \alpha_s \) against the parameter \( \kappa = g \), with color coding indicating the number of e-folds \( N_0 \).
}

       \label{fig2}
\end{figure}

\begin{figure}[htbp]
    \centering
    \includegraphics[width=0.48\linewidth,height=6.7cm]{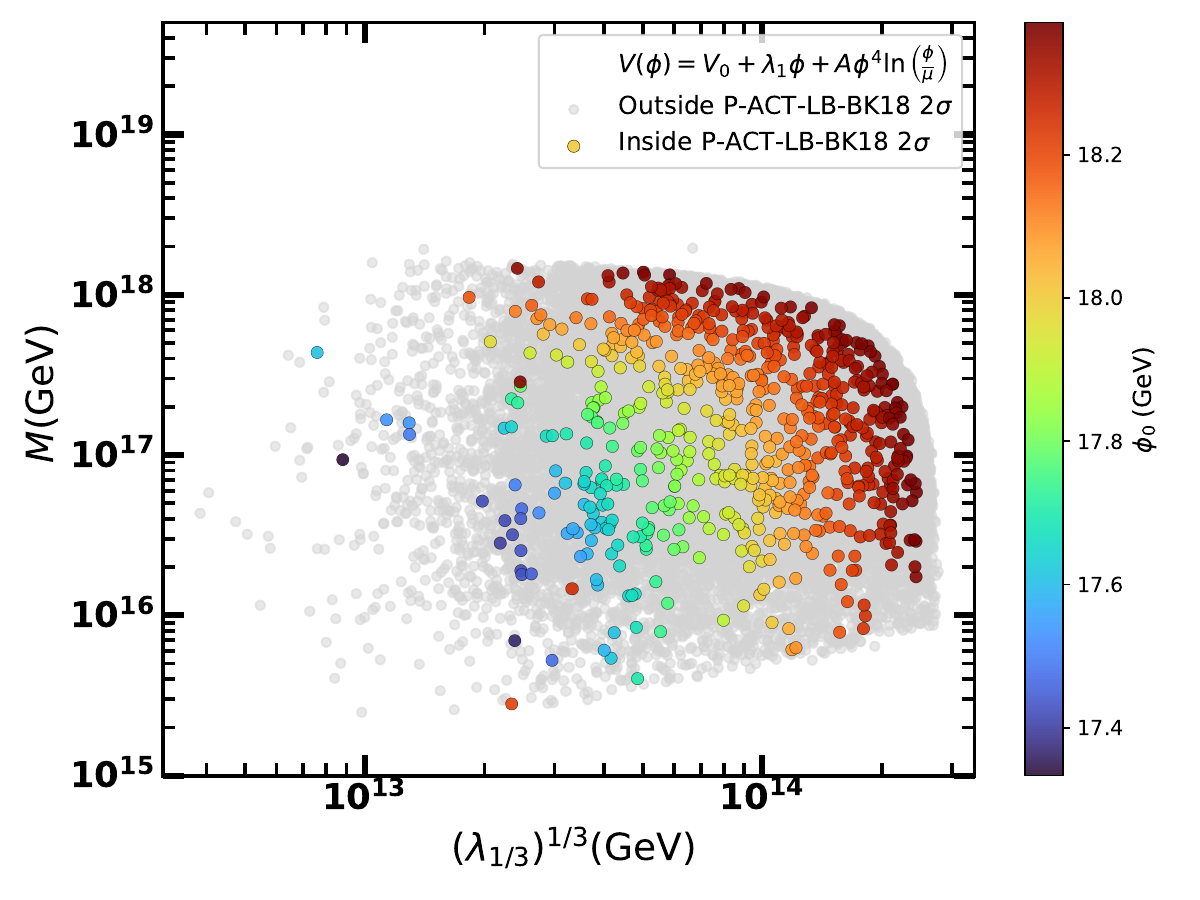}
    \quad
    \includegraphics[width=0.48\linewidth,height=6.7cm]{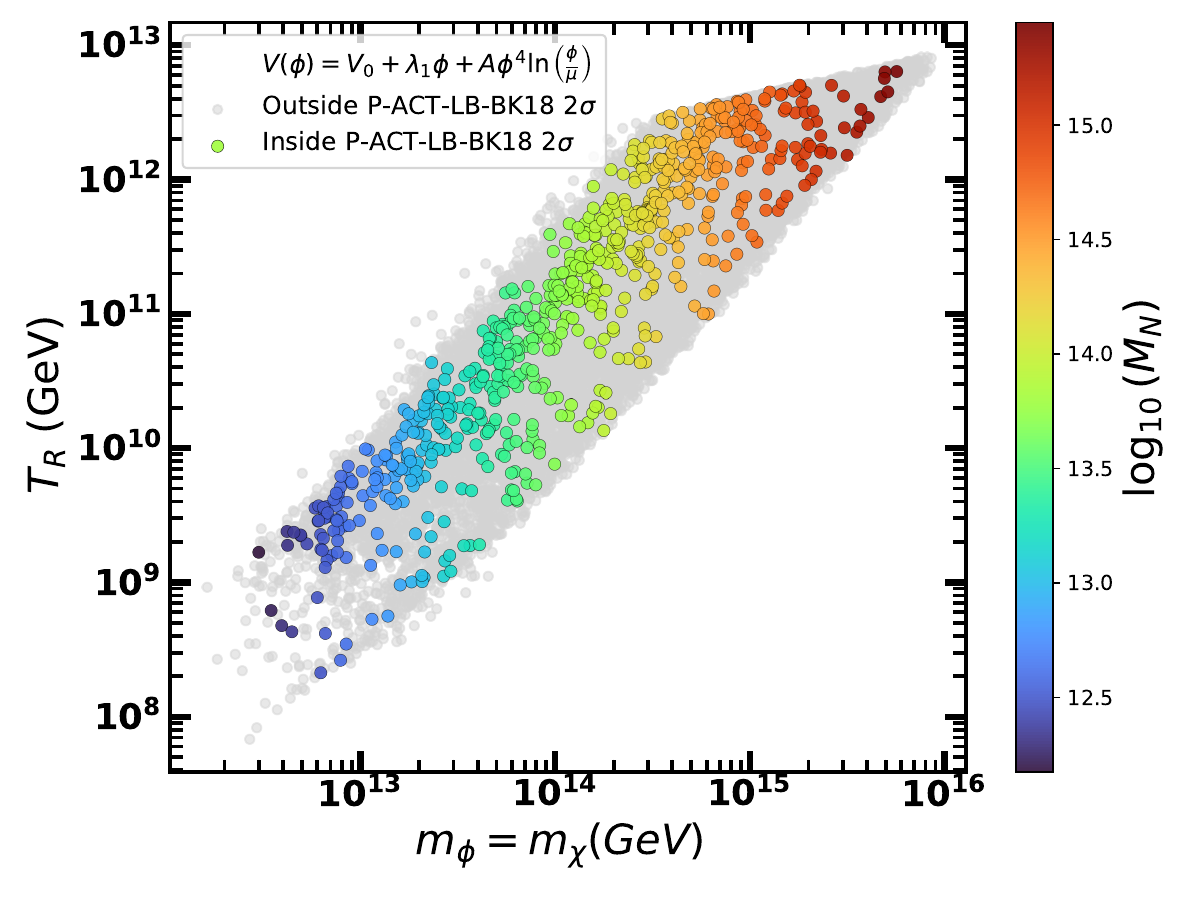}
      \quad
      \includegraphics[width=0.48\linewidth,height=6.7cm]{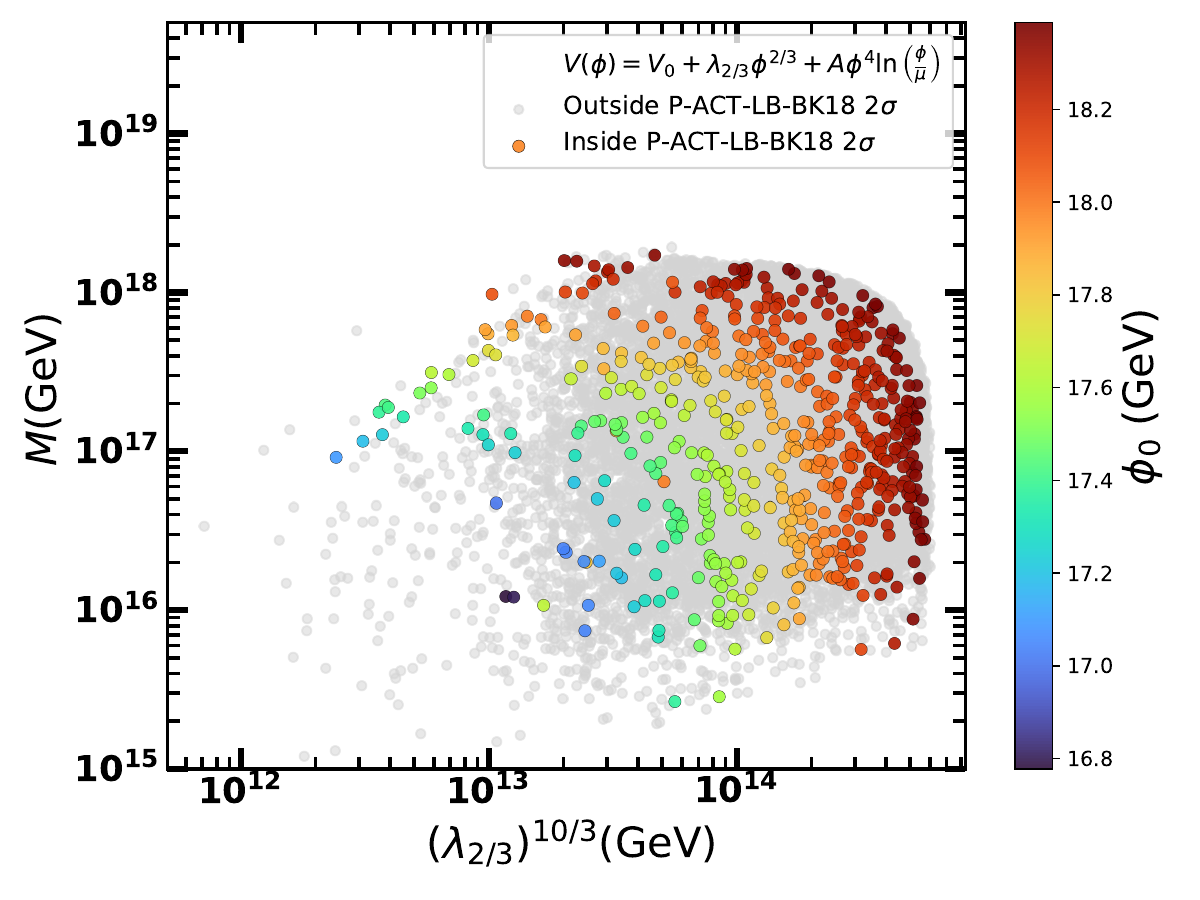}
      \quad
       \includegraphics[width=0.48\linewidth,height=6.7cm]{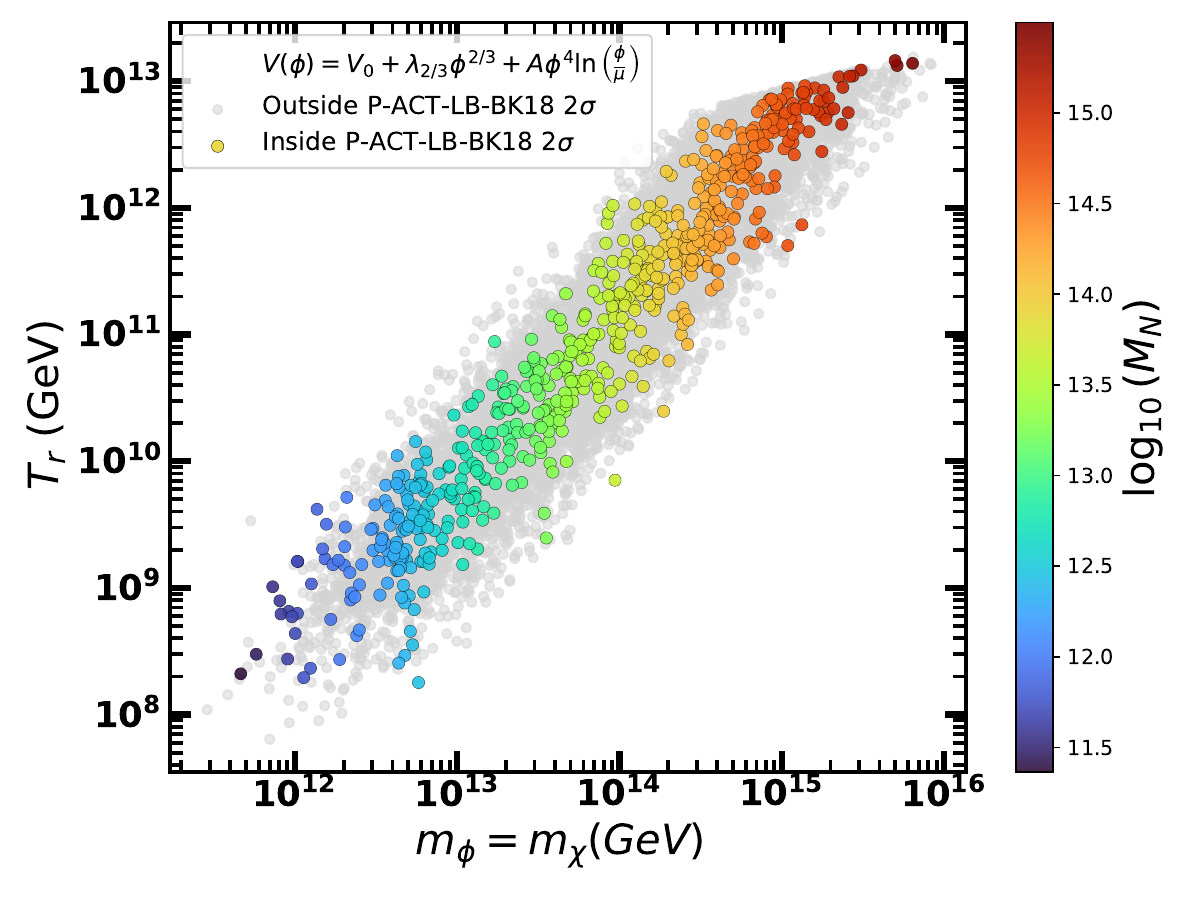}
      \quad
           \includegraphics[width=0.48\linewidth,height=6.7cm]{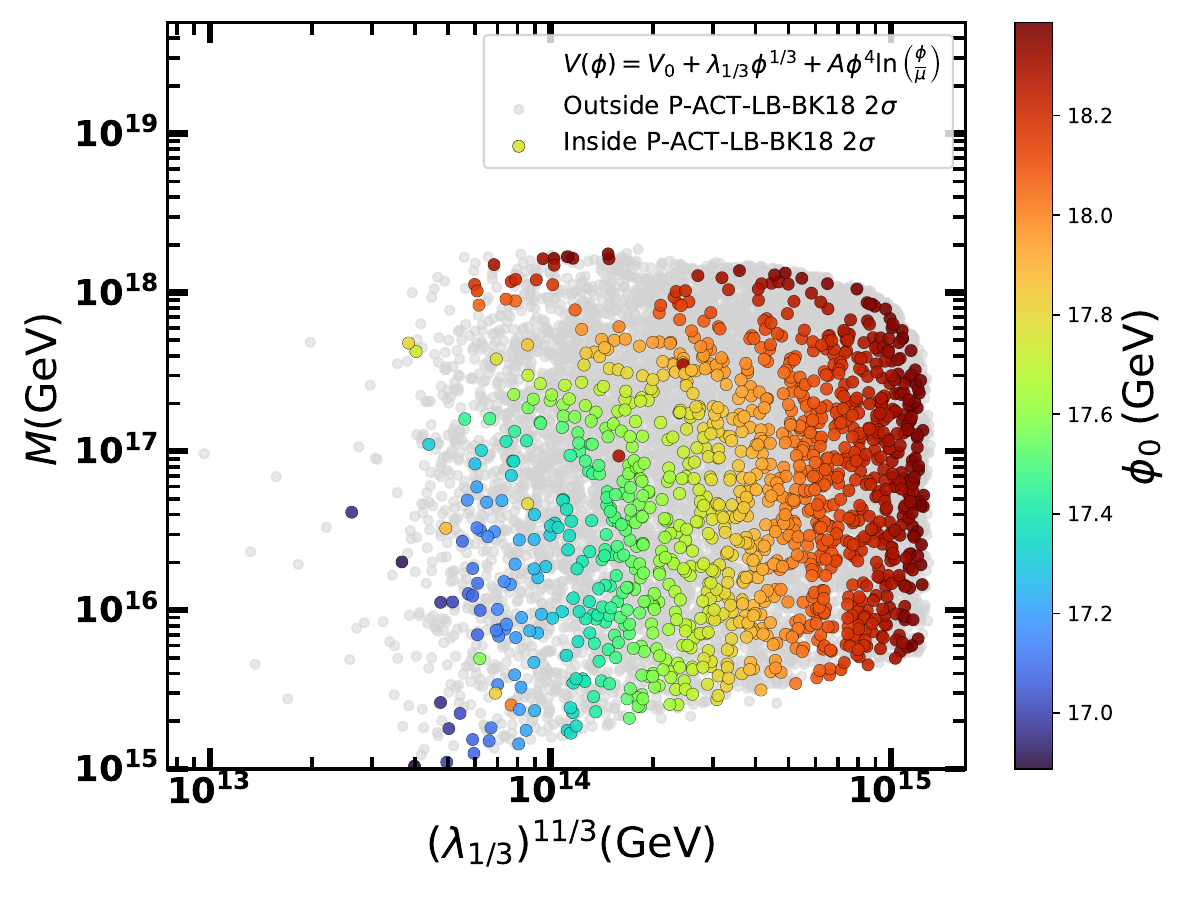}
      \quad
       \includegraphics[width=0.48\linewidth,height=6.7cm]{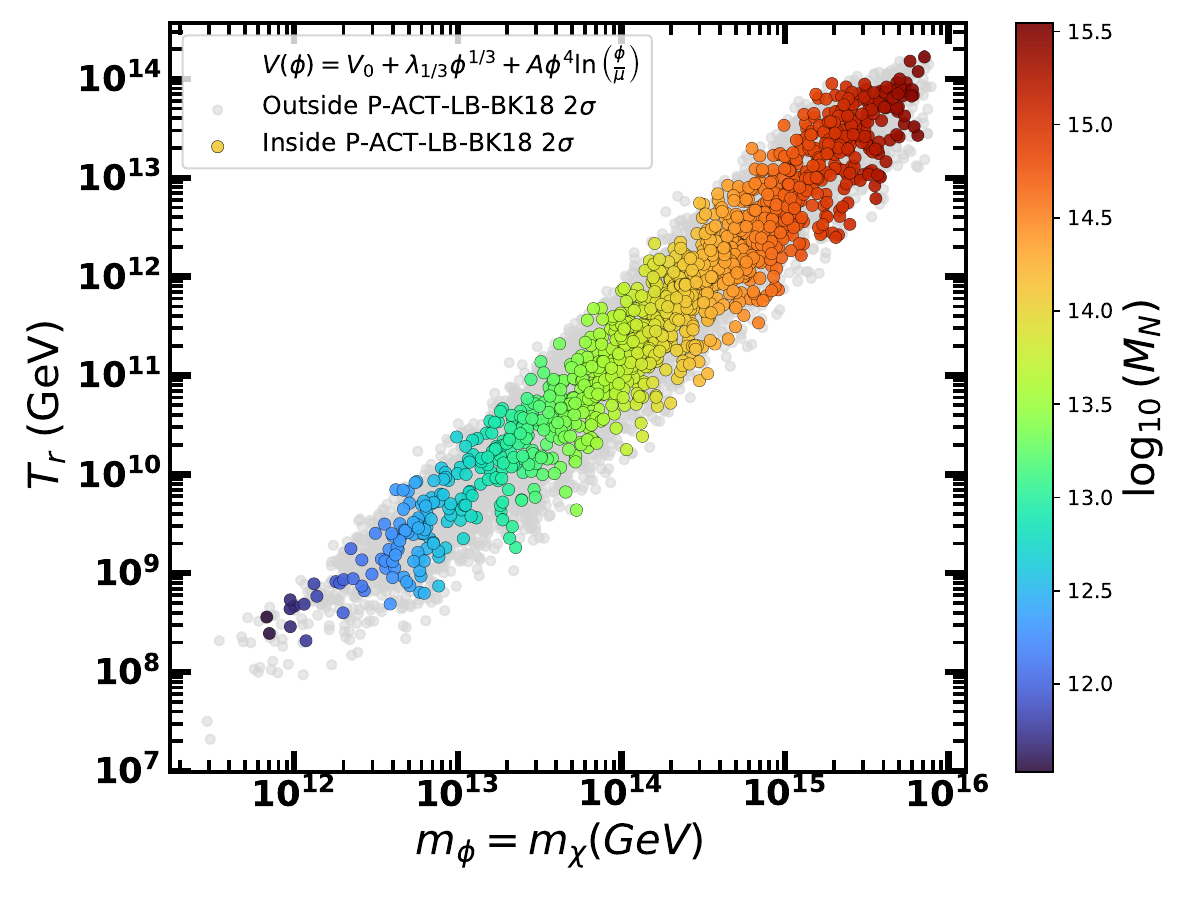}
      \quad
    \caption{
The left panel shows the coupling constant \( \lambda_p^{\frac{1}{4-p}} \) versus the symmetry breaking scale \( M \), with color coding representing the field value \( \phi_0 \). The right panel presents the inflaton mass \( m_\phi = m_{\chi} \) as a function of the reheating temperature \( T_r \), where the color coding indicates the lightest right-handed neutrino mass scale \( M_N \). All the data points shown in the rain bow color satisfy the observational constraints on the scalar spectral index \( n_s \) and tensor-to-scalar ratio \( r \) from Planck 2018, ACT DR6, and BICEP/Keck 2018 (P + ACT + LB + BK18)~\cite{AtacamaCosmologyTelescope:2025nti}.
}
\label{fig3}
\end{figure}

We present the model’s predictions for the inflationary observables in Figs.~\ref{fig2} and \ref{fig3}, considering the cases $p = 1/3$, $2/3$, and $1$. All three versions of the model predict an observable tensor-to-scalar ratio $r \lesssim 10^{-3}$, just within reach of upcoming CMB experiments such as LiteBIRD~\cite{LiteBIRD:2022cnt,LiteBIRD:2024wix}, CMB-S4~\cite{CMB-S4:2020lpa}, and the Simons Observatory~\cite{SimonsObservatory:2018koc}. For $r \gtrsim 0.01-0.001$, the field value at the pivot scale $\phi_0$ becomes trans-Planckian, consistent with the Lyth bound~\cite{Lyth:1996im}.
Successful leptogenesis is achieved by choosing the lightest right-handed neutrino mass as $M_N \simeq m_{\phi}/2$, and requiring $M_N > T_r$ to avoid thermal washout of the generated asymmetry.
The values of the symmetry-breaking scale $M$ found in our scan typically lie near the grand unification scale, $2 \times 10^{16}$~GeV, supporting the realization of this model within a GUT framework. Importantly, our viable solutions satisfy $y_\phi > g$, indicating a dominance of fermionic radiative corrections, which facilitates both reheating and non-thermal leptogenesis, features not explored in earlier works~\cite{Senoguz:2008nok,Rehman:2009wv,Rehman:2010es,Ahmed:2014cma,Bostan:2019fvk}.
Unlike radiatively corrected inflationary models with a nonminimal coupling to gravity~\cite{Okada:2010jf,Bostan:2019fvk,Ahmed:2025rrg}, the present setup avoids the theoretical frame-dependent ambiguities associated with calculations of radiative correction~\cite {Bezrukov:2013fka,Bezrukov:2009db,Bezrukov:2008ej,Hamada:2016onh}. The running of the scalar spectral index remains well within observational bounds, with $\alpha_s \lesssim 0.001$ throughout the viable parameter space.

To gain some insight into our numerical work by calculating the approximate analytical expressions for inflationary observables, we express the effective potential in terms of suitably rescaled dimensionless variables as follows:
\begin{equation}
V = V_{0} + \lambda_{p} \phi^{p} -  |A| \, \phi^{4} \ln \left(\frac{\phi}{\phi_{c}}\right) =  V_{0}\left(1 + \beta_p \, x^{p} -  \gamma  \, x^{4} \ln x \right),
\end{equation}
where $x \equiv \phi / \phi_c$ and
\begin{equation}
\beta_p \equiv \frac{\lambda_p \phi_c^p}{V_0}, \quad
\gamma \equiv \frac{|A|\phi_c^4}{V_0},
\end{equation}
are dimensionless parameters describing the relative strength of the terms in the above potential. The form of this potential with $p=2$ becomes similar to the one recently considered in the tribrid inflation framework \cite{Ahmed:2025crx}, except for the extra factor of log-term. In supersymmetry hybrid and tribrid inflation models \cite{urRehman:2006hu,Rehman:2009nq,Rehman:2010wm,Ahmad:2025mul,Antusch:2004hd,Masoud:2021prr}, these polynomial terms arise from the supergravity corrections. Note that the 1-loop corrections in supersymmetric models are approximated as $\ln(x)$ in the large field limit \cite{Dvali:1994ms,Rehman:2025fja} as compared to $x^4 \ln x$ term in non-supersymmetric models.
This important difference leads to relatively distinct predictions as derived below.

To derive the analytical expression for the inflationary observables, we first calculate the standard slow-roll parameters, which are given by
\begin{align}
\epsilon (x) &= \frac{1}{2} \left( \frac{m_P}{\phi_c} \right)^{2}  \left( \beta_p \, p \, x^{p-1} - \gamma \, x^3 ( 4 \ln x + 1 ) \right)^{2},  \\
\eta (x) &= \left( \frac{m_P}{\phi_c} \right)^{2}  \left( \beta_p \, p (p-1) x^{p-2} - \gamma \, x^2 ( 12 \ln x + 7 ) \right), \\
\zeta^2 (x) &=  \left( \frac{m_P}{\phi_c} \right)^4 
\left( 
\frac{
\left[ \beta_p p x^{p-1} - \gamma x^3 (4 \ln x + 1) \right]
\left[ \beta_p p(p-1)(p-2) x^{p-3} - \gamma x (24 \ln x + 26) \right]
}{
\left( 1 + \beta_p x^p - \gamma x^4 \ln x \right)^2
}
\right).
\end{align}
Using the expression of the amplitude of the scalar power spectrum in the slow-roll approximation given in the appendix, we obtain 
\bea
A_s &=&  \frac{1}{12 \pi^{2}} \left(\frac{V_{0}}{m_P^4} \right) \left(\frac{\phi_c^2}{m_P^2} \right) \left( \beta_p \, p \, x_0^{p-1} - \gamma \, x_0^3 ( 4 \ln x_0 + 1 ) \right)^{-2} = 2.137 \times 10^{-9},
\label{dR0}
\eea
where  $x_0 \equiv x(k_0)$ is the value of the inflaton field when the pivot scale exits the horizon.
The number of e-folds corresponding to the horizon exit of the scale $l_{0} =\frac{2 \pi}{k_{0}}$ is given by
\bea
N_{0} &=& \left( \frac{\phi_c}{m_P} \right)^2 \int_{1}^{x_0} \frac{V(x)}{V'(x)} \, dx = \left( \frac{\phi_c}{m_P} \right)^2 
\int_{1}^{x_0} 
\frac{1 }
     { \beta_p p x^{p-1} - \gamma x^3 (4 \ln x + 1) } \, dx.
\eea
To leading order, the tensor-to-scalar ratio $r$, the scalar spectral index $n_{s}$,  and the running of scalar spectral index $\alpha_s$ are given by
\bea
r &\simeq& 8 \left( \frac{m_P}{\phi_c} \right)^2 
\left( \beta_p \, p \, x_0^{p-1} - \gamma \, x_0^3 (4 \ln x_0 + 1) \right)^2 = \frac{2}{3 \pi^2 A_s}
\frac{V_0}{m_P^4},  \\
n_{s} &\simeq& 1 + 2 \eta(x_0) - \frac{3}{8} r, \quad  
\alpha_s \simeq \frac{r}{2}(n_s - 1) - 2 \zeta^2 (x_0).
\eea
Note that in our numerical computations, we have adopted next-to-leading order (NLO) slow-roll approximations for evaluating \(n_s\), \(r\), and \(A_s\), as outlined in Refs.~\cite{Stewart:1993bc, Kolb:1994ur}. We present three benchmark points, one for each model with $p=1, \, 2/3$ and $1/3$, in Table \ref{tab:benchmarks}.

Let us consider Benchmark Point 1 with $p = 1$, where the inflaton field value is near the Planck scale, $\phi_0 \simeq m_P$. For this case, our numerical analysis yields a tensor-to-scalar ratio of $r \simeq 0.001$ and a scalar spectral index $n_s = 0.971$, as presented in the first column of Table~\ref{tab:benchmarks}. The analytical expressions for the inflationary observables simplify significantly in this regime:
\bea
r &\simeq& 4 \left( \frac{m_P}{M} \right)^2 \left[ \beta_1 -  4 \gamma x_0^3 \right]^2 = \frac{2}{3\,\pi^2 A_s} \frac{\kappa^2 M^4}{m_P^4}, \quad n_s   \simeq  1 - 12 \gamma \left( \frac{m_P}{M} \right)^{2}  \left( x_0^2 \, \ln x_0   \right) , 
 \\
\alpha_s & \simeq &  \frac{r}{2}(n_s - 1) ,  \quad  N_0 \simeq \left( \frac{\phi_c}{m_P} \right)^2 \frac{x_0}{\beta} \left( 1 + \frac{\gamma}{\beta} x_0^3 \ln x_0 \right).
\eea
Using the benchmark values $M/m_P \simeq 0.01$ and $x_0 \simeq 10^2$, which yield $r \simeq 0.001$, we can extract the following parameter set:
\be
\kappa \simeq 0.08, \quad
\beta_1 \simeq 0.0002,  \quad
\gamma \simeq 5 \times 10^{-12}, \quad  \alpha_s \simeq -1.5 \times 10^{-5}, \quad N_0  \simeq 56 \, .
\ee
These estimates show excellent agreement with the numerical results, highlighting the internal consistency of our framework. Similar consistency checks can be performed for the remaining benchmark points.
Moreover, in the effective field theory limit with sub-Planckian field values ($\phi_0 < m_P$), our parameter scans confirm that combinations satisfying these analytical relations yield inflationary predictions well aligned with observational data. Notably, this model allows for a potentially measurable tensor-to-scalar ratio, making it testable in forthcoming CMB polarization experiments.

\begin{table}[t]
\centering
\renewcommand{\arraystretch}{1.2}
\setlength{\tabcolsep}{6pt}
\begin{tabular}{|c|c|c|c|}
\hline
\textbf{Parameter} & \textbf{Benchmark 1} & \textbf{Benchmark 2} & \textbf{Benchmark 3} \\
\hline
\textbf{Inflationary Potential} &
$V_0 + \lambda_1 \phi + A \phi^4 \ln\left(\frac{\phi}{\mu}\right)$ &
$V_0 + \lambda_{2/3} \phi^{2/3} + A \phi^4 \ln\left(\frac{\phi}{\mu}\right)$ &
$V_0 + \lambda_{1/3} \phi^{1/3} + A \phi^4 \ln\left(\frac{\phi}{\mu}\right)$ \\
\hline
$M$ [GeV]               & $2.1 \times 10^{16}$     & $1.9 \times 10^{16}$       & $2.1 \times 10^{16}$       \\
$\lambda_p^{\frac{1}{4-p}}$ [GeV]               & $2.1 \times 10^{14}$     & $3.01 \times 10^{15}$       & $1.01 \times 10^{15}$       \\
$\kappa = g$                & $8.75 \times 10^{-2}$     & $5.14 \times 10^{-2}$       & $5.35 \times 10^{-5}$       \\
$y_{\phi} = y_{\chi}$              & $0.14$                   & $2.29$                      & $4.47$       \\
$-A$                    & $1.4 \times 10^{-14}$    & $5.14 \times 10^{-15}$      & $2.5 \times 10^{-15}$      \\
$\phi_0$ [GeV]          & $2.1 \times 10^{18}$     & $1.35 \times 10^{18}$       & $2\times 10^{18}$       \\
$\phi_c$ [GeV]          & $2.9 \times 10^{16}$     & $2.73 \times 10^{16}$       & $2.3 \times 10^{16}$       \\
$n_s$                   & $0.9707$                  & $0.9745$                    & $0.973$                    \\
$r$                     & $1.1 \times 10^{-3}$     & $3.0 \times 10^{-4}$       & $4.7 \times 10^{-4}$       \\
$-\alpha_s$              & $1.66 \times 10^{-5}$    & $3.6 \times 10^{-6}$      & $6.2 \times 10^{-6}$      \\
$n_L/s$                 & $2.0 \times 10^{-10}$     & $2.0 \times 10^{-10}$       & $2.0 \times 10^{-10}$       \\
$T_r$ [GeV]             & $1.6 \times 10^{12}$     & $8.2 \times 10^{12}$       & $1.34 \times 10^{13}$        \\
$m_{\mathrm{inf}}$ [GeV]  & $2.6 \times 10^{15}$     & $1.4 \times 10^{15}$       & $1.6 \times 10^{15}$       \\
$\Gamma_{\phi}$ [GeV]   & $3.6 \times 10^{6}$      & $9.5 \times 10^{7}$        & $2.54 \times 10^{8}$        \\
$M_N$ [GeV]             & $1.28 \times 10^{15}$     & $7 \times 10^{14}$       & $8\times 10^{14}$       \\
$N_0$             & $57.2$                   & $58.9$                     & $57.4$                     \\
\hline
\end{tabular}
\caption{Benchmark points for inflationary potentials and reheating parameters.}
\label{tab:benchmarks}
\end{table}

\begin{figure}[htbp]
    \centering
    \includegraphics[width=0.48\linewidth,height=6.7cm]{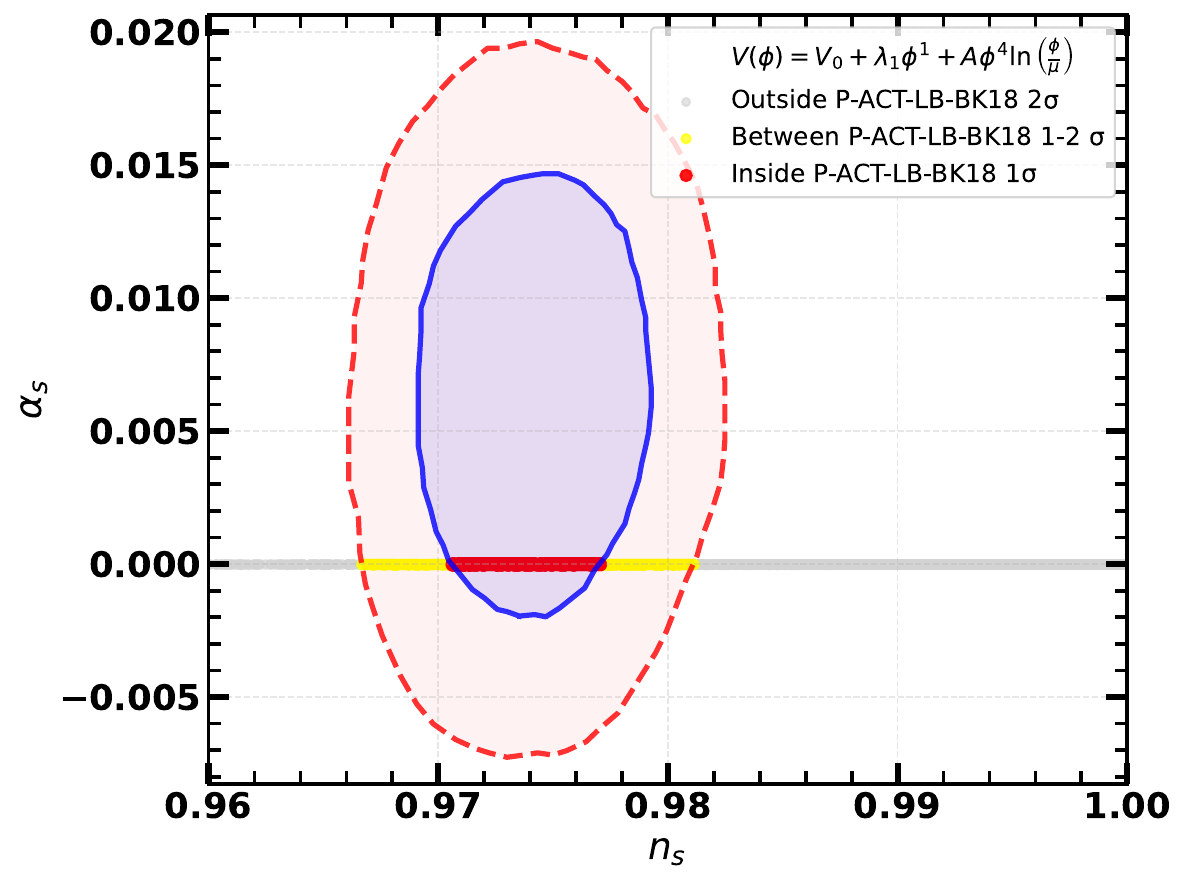}
    \quad
    \includegraphics[width=0.48\linewidth,height=6.7cm]{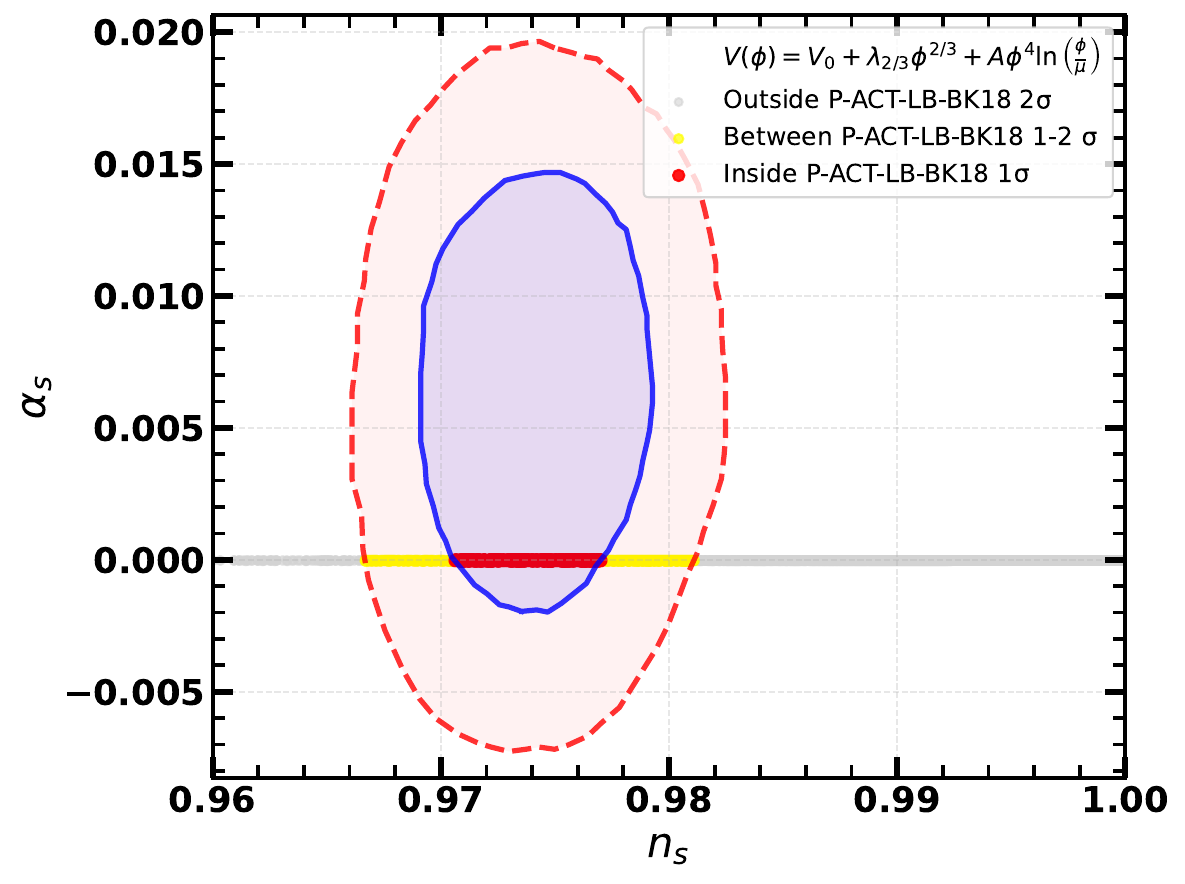}
      \quad
      \includegraphics[width=0.48\linewidth,height=6.7cm]{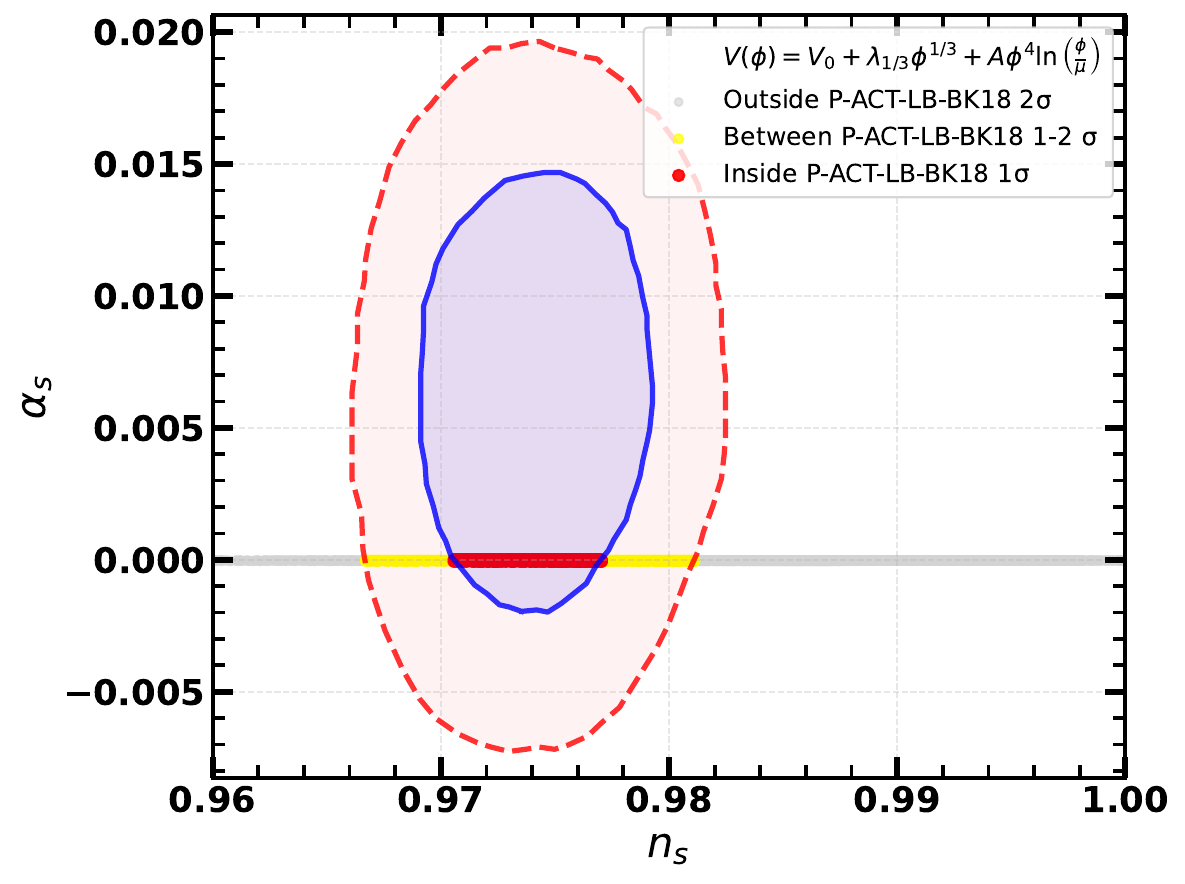}
      \quad
    \caption{
Predicted values of the scalar spectral index $n_s$ and its running $\alpha_s \equiv d n_s / d\ln k$ for radiatively corrected hybrid inflation with chaotic polynomial potentials of powers $p = 1$, $2/3$, and $1/3$ (from the upper-left to the lower panels), compared with the 1$\sigma$ and 2$\sigma$ confidence regions from the latest combined \textit{Planck} 2018 and ACT DR6 data (P–ACT–LB) \cite{AtacamaCosmologyTelescope:2025nti}.
. 
}
\label{fig4}
\end{figure}

In the present class of models, the running of the scalar spectral index is predicted to be negative but extremely small, with $|\alpha_s| < 10^{-5}$. For a negligible tensor-to-scalar ratio ($r \simeq 0$), the ACT collaboration reports a $1\sigma$ allowed range for the scalar spectral index given by
\begin{equation}
0.9707 \lesssim n_s \lesssim 0.977 .   
\end{equation}
We find that a sizable region of the model parameter space yields values of $n_s$ within this interval, as illustrated in Fig.~\ref{fig4}. Moreover, the broader $2\sigma$ P–ACT–LB range,
\begin{equation}
0.9668 \lesssim n_s \lesssim 0.9808 ,
\end{equation}
assuming $r = 0$, encompasses nearly all of our predicted points. These points with $r \lesssim 10^{-3}$ also lie within the corresponding $2\sigma$ constraints in the ($r,n_s$) plane obtained from the combined P+ACT+LB+BK18 analysis under the assumption $\alpha_s = 0$.
We further note that these constraints are expected to become less restrictive once all three CMB observables ($n_s,,\alpha_s,,r$) are allowed to vary simultaneously in fits to the ACT+DESI data.

\section{Conclusion} \label{sec7}
In this work, we have investigated a class of non-supersymmetric hybrid inflation models where the tree-level inflaton potential follows a chaotic, polynomial-like form,
$$
V(\phi) = V_0 + \lambda_p \phi^p.
$$
as given in Eq.~\ref{eq:01}. Such models, in both the chaotic limit ($V_0 \ll \lambda_p \phi^p$) and the hybrid limit ($V_0 \gg \lambda_p \phi^p$), typically yield predictions that are in tension with recent cosmic microwave background (CMB) observations from Planck and the Atacama Cosmology Telescope (ACT). The chaotic regime tends to produce an excessively large tensor-to-scalar ratio $r$, while the hybrid regime often predicts a blue-tilted scalar spectral index $n_s > 1$, both of which are disfavored by data.

We have shown that the inclusion of one-loop radiative corrections, arising naturally from inflaton couplings to other fields required for reheating, can significantly improve the observational viability of these models. The corrected potential takes the form,
$$
V(\phi) = V_0 + \lambda_p \phi^p + A \, \phi^4 \ln \phi,
$$
as given in Eq.~\ref{eq:02}, where $A$ encodes the strength of loop effects. For negative values of $A$, as expected from fermionic radiative contributions, the logarithmic term flattens the potential in the large-field regime. This leads to a red-tilted scalar spectral index ($n_s < 1$) and a suppressed tensor-to-scalar ratio $r$, both in better agreement with observational constraints. Notably, these improvements persist even for sub-Planckian field excursions, preserving theoretical control.

In the chaotic limit, we find that the model with $p = 1/3$ yields predictions well within the 1$\sigma$ region of current $r$–$n_s$ bounds, while $p = 2/3$ falls within the 2$\sigma$ contour. The case $p = 1$, despite achieving a lower $r$ and redder $n_s$, remains outside the 2$\sigma$ region. However, in the hybrid limit, all three cases, $p = 1,\, 2/3,\, 1/3$, predict observables fully consistent with the most recent Planck+ACT data at the 1$\sigma$ level.
Moreover, each scenario allows for successful reheating and accommodates non-thermal leptogenesis, providing a unified framework that links inflationary dynamics with the generation of the matter–antimatter asymmetry.
Overall, our analysis emphasizes the critical role of quantum corrections in inflationary model building and demonstrates how they can broaden the phenomenological viability of non-supersymmetric hybrid inflation frameworks.

\section{Appendix: Slow-roll Definitions}\label{sec8}
The slow-roll parameters are defined as
\begin{align}
\epsilon(\phi) &\equiv \frac{m_P^2}{2} \left( \frac{V'(\phi)}{V(\phi)} \right)^2, \quad \eta(\phi) \equiv m_P^2 \frac{V''(\phi)}{V(\phi)}, \\
\zeta^2(\phi) &\equiv m_P^4 \frac{V'(\phi) V'''(\phi)}{V^2(\phi)},
\end{align}
where primes denote derivatives with respect to the inflaton field \(\phi\), and \(m_P = (8 \pi G)^{-1/2}\) is the reduced Planck mass. These parameters are assumed to be small during inflation: \((\epsilon, |\eta|, \zeta^2) \ll 1\).
Unlike chaotic inflation, in hybrid scenarios inflation ends when the inflaton reaches a critical value \(\phi_c\), triggering a tachyonic instability in the waterfall field and causing a rapid waterfall transition.

The total number of e-folds from the time when the pivot scale exits the horizon until the end of inflation is given by:
\begin{equation}
N_0 \simeq \frac{1}{m_P^2} \int_{\phi_{\text{end}}}^{\phi_0} \frac{V}{V'} \, d\phi,
\end{equation}
where \(\phi_0 \equiv \phi(k_0)\) is the field value at horizon exit and \(\phi_{\text{end}}\) corresponds to the end of inflation, typically defined by the condition \(\max[\epsilon(\phi_{\text{end}}), |\eta(\phi_{\text{end}})|, |\zeta^2(\phi_{\text{end}})|] = 1\) or \(\phi_{\text{end}} = \phi_c\). The required number of e-folds is generally in the range \(50 \lesssim N \lesssim 60\), depending on the post-inflationary thermal history and the reheating temperature.

The amplitude of the curvature perturbation $A_s$ evaluated at the pivot scale $k_0 = 0.05\, \mathrm{Mpc}^{-1}$ is given by:
\begin{equation}
A_s(k_0) \simeq \frac{1}{12 \pi^2 m_P^6} \frac{V^3}{|V'|^2} \Big|_{\phi = \phi_0}.
\end{equation}
According to the latest Planck results \cite{Planck:2018vyg, Planck:2018jri}, this amplitude is constrained to:
\begin{equation}
A_s(k_0) \simeq 2.215 \times 10^{-9}.
\end{equation}

The key inflationary observables, the scalar spectral index \(n_s\), the tensor-to-scalar ratio \(r\), and the running of the spectral index \(\alpha_s \equiv \frac{d n_s}{d \ln k}\), are given within the slow-roll approximation by:
\begin{align}
n_s &\simeq 1 - 6 \, \epsilon(\phi_0) + 2 \, \eta(\phi_0), \quad r \simeq 16 \, \epsilon(\phi_0), \\
\alpha_s &\simeq 16 \, \epsilon(\phi_0) \, \eta(\phi_0) - 24 \, \epsilon^2(\phi_0) - 2 \, \zeta^2(\phi_0).
\end{align}
For improved accuracy, our numerical evaluations incorporate the first-order corrections to the slow-roll expansion as derived in~\cite{Stewart:1993bc}, particularly for the scalar spectral index \(n_s\), the tensor-to-scalar ratio \(r\), its running \(\alpha_s\), and the power spectrum amplitude \(A_s\).

\section*{DATA AVAILABILITY}
The data that support the findings of this article are openly available at Ref.~\cite{datta}.

\bibliographystyle{apsrev4-1}

\bibliographystyle{unsrt}  
\bibliography{References}  

\end{document}